\documentclass[preprintnumbers, floatfix, letterpaper, twocolumn,aps,prd,epsfig,nofootinbib,natbib,longbibliography]{revtex4-1}

\usepackage{graphicx}
\usepackage{epstopdf}
\usepackage{latexsym}
\usepackage{amssymb}
\usepackage{amsmath}
\usepackage{color}
\usepackage{mathrsfs}

\usepackage[center]{subfigure}

\begin{document}

 \newcommand{\bq}{\begin{equation}}
 \newcommand{\eq}{\end{equation}}
 \newcommand{\bqn}{\begin{eqnarray}}
 \newcommand{\eqn}{\end{eqnarray}}
 \newcommand{\nb}{\nonumber}
 \newcommand{\lb}{\label}
\newcommand{\PRL}{Phys. Rev. Lett.}
\newcommand{\PL}{Phys. Lett.}
\newcommand{\PR}{Phys. Rev.}
\newcommand{\CQG}{Class. Quantum Grav.}

%
\title{Towards Cosmological Dynamics  from  Loop Quantum Gravity }

\author{Bao-Fei Li $^{1,2}$}
\email{ Bao-Fei$\_$Li@baylor.edu}
\author{Parampreet Singh$^3$}
\email{psingh@phys.lsu.edu}
\author{Anzhong Wang$^{1, 2}$\footnote{The corresponding author}}
\email{Anzhong$\_$Wang@baylor.edu}
\affiliation{$^{1}$Institute for Advanced Physics $\&$ Mathematics,
Zhejiang University of Technology, Hangzhou, 310032, China\\
$^2$GCAP-CASPER, Department of Physics, Baylor University, Waco, TX, 76798-7316, USA\\
$^3$ Department of Physics and Astronomy, $\&$ Center for Computation and Technology, Louisiana State University, Baton Rouge, LA 70803, USA}

\date{\today}

\begin{abstract}

We present a systematic study of the cosmological dynamics resulting from an effective Hamiltonian, recently derived in loop quantum gravity 
using Thiemann's regularization and earlier obtained in loop quantum cosmology (LQC) by keeping the Lorentzian term explicit in the Hamiltonian 
constraint. We show that quantum geometric effects result in higher than quadratic corrections in energy density in comparison to LQC causing a 
non-singular bounce. Dynamics can be described by the Hamilton's or the Friedmann-Raychaudhuri equations, but the map between the two descriptions 
is not one-to-one. A careful analysis resolves the tension on symmetric versus asymmetric bounce in this model, showing that the bounce must be 
asymmetric and symmetric bounce is physically inconsistent, in contrast to the standard LQC. In addition, the current observations only allow a scenario 
where the pre-bounce branch is asymptotically de Sitter, similar to a quantization of the Schwarzschild interior in LQC, and the post-bounce branch 
yields the classical general relativity. For a quadratic potential, we find that a slow-roll inflation generically happens after the bounce, which is quite 
similar to what happens in LQC.

\end{abstract}
\maketitle

\section{Introduction}
\label{Intro}
\renewcommand{\theequation}{1.\arabic{equation}}\setcounter{equation}{0}

In the regime where spacetime curvature reaches Planck scale and Einstein's 
theory of general relativity (GR) breaks down, it has been long expected that a quantum theory of gravity will provide insights on the resolution of the big bang singularity. 
How  such a  resolution of cosmological and black hole singularities takes place is a  
fundamental question which any candidate theory of quantum gravity must answer. Loop quantum gravity (LQG) is one of the main candidate theories of quantum gravity, with a key prediction that classical 
differential geometry at small spacetime curvatures is replaced by a discrete quantum geometry at the Planck scale \cite{reviewlqg}. In the last decade, various cosmological and black hole spacetimes have been studied using techniques of 
LQG in loop quantum cosmology (LQC) where a classical symmetry reduction is performed for  spacetimes before quantization. The main result of LQC is the existence of a big bounce when the energy density of matter reaches a maximum value 
determined by the underlying quantum geometry \cite{aps,aps3,slqc}. There have been many extensions, robustness checks and applications of these results in cosmological and black hole settings in the last decade (for reviews see \cite{review1,review2}). 
Physical implications have been extensively studied using an effective spacetime 
description of the quantum spacetime derived using coherent states \cite{vt}. On the question of big bang and other singularities, effective dynamics of  LQC gives a definitive answer. It predicts a generic resolution of all strong curvature singularities in 
isotropic and anisotropic spacetimes \cite{generic}. And, there are many investigations linking the Planck scale physics of LQC with cosmological observations (for reviews of different approaches see for eg. \cite{cmb} and references there in).

Despite a wealth of results on the singularity resolution and phenomenology of the very early universe obtained in LQC, an important issue which has remained open is its connection with LQG (see for example \cite{engle} for discussions). One of the key questions is whether the cosmological 
dynamics from LQG bears a close similarity to the one uncovered in LQC. Here we should note that the quantization procedure used in LQC owing to symmetry reduction before quantization can result in a different Hamiltonian constraint than the choice made in LQG, hence, resulting in different Planck scale physics. Though the question of ambiguities in obtaining the Hamiltonian and the resulting dynamics in LQG is far from settled, there have been ongoing attempts to extract hints based on some rigorous proposals by Thiemann. In particular, in the cosmological setting, one of the first attempts to understand this issue was made by Yang, Ding and Ma \cite{YDM09}, by considering a Hamiltonian constraint in LQC which is closer to the actual construction in LQG using Thiemann's regularization of the Hamiltonian constraint \cite{thiemann}. In this work, unlike the quantum Hamiltonian constraint in LQC which turns out to be a second order quantum difference equation, a fourth order quantum difference equation was obtained. Notably, a similar result on the form of the quantum difference equation was obtained earlier by Bojowald \cite{martin}, albeit for an early quantization of LQC which is not physically viable \cite{cs08,aps3}. The difference in the Hamiltonian constraints of these studies from the standard Hamiltonian constraint in LQC stems from the following. The Hamiltonian constraint in LQG is composed of a Euclidean and a Lorentzian term which 
are proportional to each other in the absence of spatial curvature. One can then rewrite the Lorentzian term in the same form as the Euclidean term before quantization, resulting in the standard Hamiltonian constraint in LQC \cite{abl}. 
This strategy does not work in LQG because the Lorentzian term needs to be quantized in a different way than the Euclidean term, leading to a difference in the quantum Hamiltonian constraint for the spatially flat cosmological spacetime. It is to be noted that even though the quantum Hamiltonian 
constraints turned out to be different, the authors of \cite{YDM09} found that there were only a little qualitative difference from LQC in the resulting physics. The only notable change which was found was in the value of energy density at the bounce. The pre-bounce and post-bounce phases were found to be symmetric and resulting in classical small curvature general relativistic spacetime at large volumes, in a close agreement with the spatially flat model in LQC \cite{aps3}.

In a recent study, Dapor and Liegener \cite{DL17} obtained expectation values of the Hamiltonian operator in LQG using complexifier coherent states \cite{states}, adapted to the spatially flat FLRW cosmological spacetime.  Their procedure, using non-graph changing regularization of the Hamiltonian \cite{thiemann}, resulted in an effective Hamiltonian which to the leading order agrees with the one in  \cite{YDM09}. Strictly speaking, the work in \cite{DL17} treated edge length $\mu$ to construct loops as a free parameter, but in Ref. \cite{YDM09} it was considered as a specific triad dependent function, the so-called $\bar \mu$ scheme \cite{aps3},  which in LQC is known to be the only possible choice resulting in physics independent from underlying fiducial structures used during quantization and yielding a consistent infra-red behavior for all matter obeying the weak energy condition \cite{cs08}. In a later work, results of \cite{DL17} have been extended to the latter scheme \cite{adlp}. In the following we will consider the effective Hamiltonians in both  approaches in the $\bar \mu$ scheme. With this note, interestingly, but in contrast to the results in  \cite{YDM09}, the authors of  \cite{DL17} reported that though the dynamics agrees with LQC on one side of the bounce, it does not agree on the other side. The bounce in spatially flat isotropic spacetime was found to be asymmetric, usually a hallmark of anisotropic and black hole spacetimes in LQC \cite{kasner1,djs}. In particular, it turns out that on one side of the bounce a quantum spacetime with an effective positive cosmological constant emerges \cite{adlp}. It is interesting to note that a similar result with an emergent cosmological constant was earlier found in LQC for a quantization of the interior of  the Schwarzschild black hole \cite{djs}.

As stressed above, due to the ambiguities in extracting Hamiltonian dynamics in LQG it is not at all clear whether the above results can be considered to accurately capture the true cosmological dynamics in LQG even to the leading order. Nevertheless, they result from a specific and a rigorous regularization of the Hamiltonian constraint in LQG and  can certainly be considered as one candidate for possible dynamics in the cosmological sector. With this important caveat, the above results confirm that the big bounce first found in LQC holds at least for a particular regularization of Hamiltonian in LQG. And it points to a qualitative difference on one side of the bounce which as we argue in this paper must be pre-bounce to our expanding cosmological spacetime. However, if we consider the results of \cite{YDM09} and \cite{DL17} as potentially capturing LQG cosmological dynamics, we are immediately led to a puzzle. The puzzling aspect of these results is  the striking difference in the nature of pre-bounce 
universe even though both  works  start from the same effective Hamiltonian. There is tension whether the pre-bounce scenario is symmetric as predicted in spatially flat isotropic LQC \cite{YDM09} or asymmetric \cite{DL17}.  Partly this disagreement is tied to a rather preliminary analysis of the effective dynamics, resulting from the effective Hamiltonian in LQG and in LQC where Lorentzian term is treated independent of the Euclidean term. In particular, neither the modified set of the Friedmann and Raychaudhuri  (FR) equations are  known so far, nor one understands the underlying conditions to get a complete physical solution resulting from the effective Hamiltonian. A rigorous understanding of the effective dynamics is needed to extract the cosmological dynamics from LQG.

The goal of this paper is to find the modified dynamical equations in LQG starting from the effective Hamiltonian found in \cite{DL17},  which coincides with the one in LQC with a Lorentzian term regularized via Thiemann's procedure \cite{YDM09}. Since their effective Hamiltonian are the same, we refer to both as the effective dynamics resulting from LQG to contrast the results with the standard LQC. We emphasize that even though we label these works as ``LQG cosmology'' to distinguish them from LQC, it does not discount the important caveat of various ambiguities in  writing the Hamiltonian in LQG discussed above. In fact, it should not be surprising if in the future the true LQG cosmological dynamics which emerges after settling various ambiguities is different from the proposal in \cite{DL17}. In this sense, our investigation should be viewed as a modest attempt to gain insights on the above proposals which take us towards the cosmological dynamics of LQG. 

Starting from the effective Hamiltonian, using the Hamilton's equations, we derive the modified FR equations in terms of the polymerized loop variables. The vanishing of the Hamiltonian constraint results in an expression for the energy density which has two roots. We show that there exist two distinct branches which need to be carefully taken into consideration in order to extract cosmological dynamics.  We find the modified FR equations which turn out to be different from their counterparts in LQC. Notably,
 in LQC the modification to the classical FR equations is via quadratic powers in 
energy density \cite{ps06,aps3}. In contrast, in the LQG cosmology we obtain higher order modifications. In LQC, the pre-bounce and post-bounce phases are governed by only one set of the modified FR equations when expressed in terms of energy density and pressure of the matter content.  In contrast, in the cosmological dynamics of LQG, the modified FR equations are different in pre-bounce and post-bounce phases when expressed in terms of energy density and pressure, coinciding only at the bounce when energy density reaches a maximum value $\rho_c$. As in LQC, the existence of $\rho_c$ is tied to the non-zero minimum area eigenvalue in LQG. Further, the maximum energy density $\rho_c$ is roughly four times smaller when compared to the bounce density in LQC. 

After finding the dynamical equations we obtain numerical solutions in two different ways. First, by using the Hamilton's equations and the second by using modified FR equations. The results from these exercises  must be identical but it turns out that because of the two branches and different sets of the modified FR equations in the pre- and post-bounce phases, one needs to be careful 
in the numerical analysis based on the latter. A naive analysis without carefully considering different roots of energy density and different sets of the FR equations needed for a complete evolution can lead one to a ``symmetric bounce,'' which on examining carefully turns out to be unphysical,  because it results from using the modified FR equations beyond their domains of validity. Or in other words, the Hamiltonian constraint is no longer satisfied for a symmetric bounce.   On the other hand, if one carefully takes into  account different roots in the modified FR equations along with their domains of validity, one obtains a picture consistent with that obtained from the Hamilton's equations. We perform numerical simulations considering a massless scalar as well as a quadratic inflationary potential. We find that the post-bounce phase is in a good agreement with LQC, where as the pre-bounce phase leads to a de Sitter regime with a quantum gravitational cosmological constant. Our results demonstrate that an inflationary phase is compatible with the cosmological dynamics in LQG. A universe starting from a de Sitter phase in the pre-bounce regime undergoes a non-singular evolution to an inflating regime in the post-bounce regime, followed by a phase of reheating.

This paper is organized as follows. In Sec. II we start with the effective Hamiltonian found using complexifier coherent states pertaining to a cosmological spacetime in LQG \cite{DL17}. The same effective Hamiltonian was found earlier in LQC using the Hamiltonian constraint with an explicit Lorentzian term \cite{YDM09}. Using the Hamiltonian formalism, we find the Hamilton's equations and obtain the modified FR ones. After studying various details of these equations we perform numerical simulations for the massless scalar field using Hamilton's equations in Sec. III, and using the modified FR equations in Sec. IV. Various subtleties are discussed and clarified in these sections. In particular, we show that a careful analysis leads to the same results. And if one ignores some subtleties, it is easy to get an unphysical symmetric bounce. In Sec. V we study the cosmological dynamics in LQG with an inflationary potential. We summarize our results in Sec. V.

\section{Effective Hamiltonian and the Modified Planck Scale Dynamics }
\label{Section2}
\renewcommand{\theequation}{2.\arabic{equation}}\setcounter{equation}{0}

In LQG, which uses Dirac's method of quantization of constraints, the elementary classical phase space variables are the SU(2) Ashtekar-Barbero connection $A^i_a$ and  the conjugate triad $E^a_i$. 
In the cosmological setting, which is homogeneous and isotropic, the only relevant constraint is the Hamiltonian constraint which for the gravitational 
sector is a sum of the Euclidean and Lorentzian parts, given by
\bq
\mathcal{C}_{\mathrm{grav}} = {\cal C}_{\mathrm{grav}}^{(E)} - (1 + \gamma^2) {\cal C}_{\mathrm{grav}}^{(L)}, 
\eq
where for the lapse chosen to be unity, the Euclidean part is 
\bq
\mathcal{C}_{\mathrm{grav}}^{(E)} = \frac{1}{16 \pi G} \int \mathrm{d}^3 x \, \epsilon_{ijk} F^i_{ab} \frac{E^{aj} E^{bk}}{|\mathrm{det}(q)|}, 
\eq
and the Lorentzian part is 
\bq
\mathcal{C}_{\mathrm{grav}}^{(L)} =  \frac{1}{8 \pi G} \int \mathrm{d}^3 x \,   K^j_{[a} K^k_{b]}   \frac{E^{aj} E^{bk}}{|\mathrm{det}(q)|}   ~.
\eq
Here $K^i_a$ captures the extrinsic curvature,  $\gamma$ is the Barbero-Immirzi parameter whose value is set to $\gamma \approx 0.2375$ using 
black hole thermodynamics in LQG \cite{Mei04}, and $|\mathrm{det}(q)|$ is the determinant of the spatial metric compatible with the triads.

As mentioned earlier, if the cosmological spacetime is spatially flat, then the Euclidean and Lorentzian terms are proportional to each other. One can then rewrite the classical 
constraint entirely in terms of the Ashtekar-Barbero connection $A^i_a$, which has been the usual strategy in LQC \cite{abl,aps,aps3}. Expressing the resulting expression in terms of holonomies of the connection one is 
led to a second order quantum difference equation with uniform discreteness in volume after quantization \cite{aps3}. The quantum evolution 
results in a big bounce, a result confirmed through extensive numerical simulations \cite{aps3,review2,numerics} and an exactly solvable model \cite{slqc}.  Using coherent states, an effective Hamiltonian can be 
derived which results in a $\rho^2$ modification to the FR equations \cite{ps06,aps3}. The validity of the effective dynamics has been rigorously tested using a wide variety of quantum states \cite{numerics}. Our analysis in this paper will be based on the assumption that the effective Hamiltonian is valid for the entire evolution and provides a good approximation to the underlying quantum dynamics in both  LQC and LQG.

As we emphasized before,  it is not necessary to express the Hamiltonian constraint such as to eliminate the Lorentzian term explicitly in LQC. And if this term is kept explicitly we are led to a different effective Hamiltonian using either the analysis of including  the Lorentzian term in LQC \cite{YDM09}, or the computation of the expectation values from the Hamiltonian in LQG using coherent states satisfying symmetries of the 
spatially flat FLRW spacetime \cite{DL17}. The resulting effective Hamiltonian is:
 \bqn
\lb{Ha}
\mathcal {H}&=& \frac{3}{8\pi G}\Big\{\frac{p^{1/2}\sin^2(\bar \mu c)}{\bar \mu^2}-\frac{p^{1/2}(\gamma^2+1)\sin^2(2\bar \mu c)}{4\bar \mu^2 \gamma^2}\Big\}\nb\\
&& ~~~~~~~+\mathcal{H}_M,
\eqn
in terms of the loop variables $c$ and $p$ which capture the connection and the triad in the homogeneous and isotropic setting and satisfy $\{c,p\}={8\pi G\gamma}/{3}$.
Here $\bar \mu\equiv \sqrt{\Delta/p}$ \cite{aps3}, with $\Delta = \lambda^2 = 4\sqrt{3}\pi\gamma \ell_{\mathrm{Planck}}^2$ being the smallest nonzero
eigenvalue of the area operator in loop quantum gravity, and $\mathcal{H}_M$ represents the Hamiltonian of the matter sector. Note that in the above effective Hamiltonian we have chosen a positive 
orientation of the triads, and ignored the modifications coming from expressing the inverse triads in terms of holonomies, which in LQC have been shown to play negligible role for spatially flat models \cite{aps3}.
 
It is convenient to adopt the following conjugate variables, first introduced in Ref. \cite{slqc}:  $b=c/p^{1/2}$ and $v=p^{3/2}$, which satisfy $\{b,v\}=4\pi G\gamma$. Here $v = v_o a^3$,
 where $v_o$ is the volume of fiducial cell in $\mathbb{R}^3$ spatial manifold, and $a$ is the scale factor of the universe. Thus, $v$ captures the physical volume of the universe. On the other hand, in the classical theory 
$b$ is proportional to the Hubble rate via the classical dynamical equations. The effective Hamiltonian then becomes,
\bqn
\lb{ham}
\mathcal {H}&=&\frac{3v}{8\pi G\lambda^2}\Big\{\sin^2(\lambda b)-\frac{(\gamma^2+1)\sin^2(2\lambda b)}{4\gamma^2}\Big\}\nb\\
&& ~~~~~~~~~ +\mathcal{H}_M ~.
\eqn
Hence,  the Hamilton's equations for the basic variables $b$ and $v$ take the  form:
\bqn
\lb{eomA}
\dot v&=&\Big\{v, \mathcal H\Big\}=\frac{3v\sin(2\lambda b)}{2\gamma \lambda}\Big\{(\gamma^2+1)\cos(2\lambda b)-\gamma^2\Big\}, \nb\\
\\
\lb{eomB}
\dot b&=&\Big\{b, \mathcal H\Big\}=\frac{3\sin^2(\lambda b)}{2\gamma \lambda^2}\Big\{\gamma^2\sin^2(\lambda b)-\cos^2(\lambda b)\Big\}\nb\\
&& ~~~~~~~~~~~~~~ -4\pi G\gamma P,
\eqn
where $P$ represents the pressure defined by $P\equiv -{\partial \mathcal{H}_M}/{\partial v}$.  Once the matter Hamiltonian $\mathcal{H}_M$ is specified, together with the
Hamiltonian constraint,
\bq
\lb{HCD}
\mathcal{C} \approx 0,
\eq
where ${\cal C} = 16 \pi G{\cal H}$,  Eqs. (\ref{eomA}) and (\ref{eomB}) will uniquely determine the evolution of the universe. 
 
To write them in the form of the FR equations, 
 we first note that  from Eq.(\ref{eomA}) we obtain the Hubble parameter $H\equiv {\dot a }/{a}={\dot v }/{3v}$, 
\bqn
\lb{1.1}
H^2 = \frac{\sin^2(2\lambda b)}{4\lambda^2 \gamma^2}\Big\{\gamma^2-(\gamma^2+1)\cos(2\lambda b)\Big\}^2.
\eqn
In addition, the acceleration of $a$ using time derivative of (\ref{eomA}) is given by
\bq
\lb{1.1b}
\frac{\ddot a}{a}=H^2+\frac{\dot b}{\gamma}\Big\{(\gamma^2+1)\cos(4\lambda b)-\gamma^2\cos(2\lambda b)\Big\}.
\eq
Further, it can be shown that
\bq
\lb{1.2}
\dot H=\frac{\dot b}{\gamma}\Big\{(\gamma^2+1)\cos(4\lambda b)-\gamma^2\cos(2\lambda b)\Big\}.
\eq

\begin{figure}
{
\includegraphics[width=8cm]{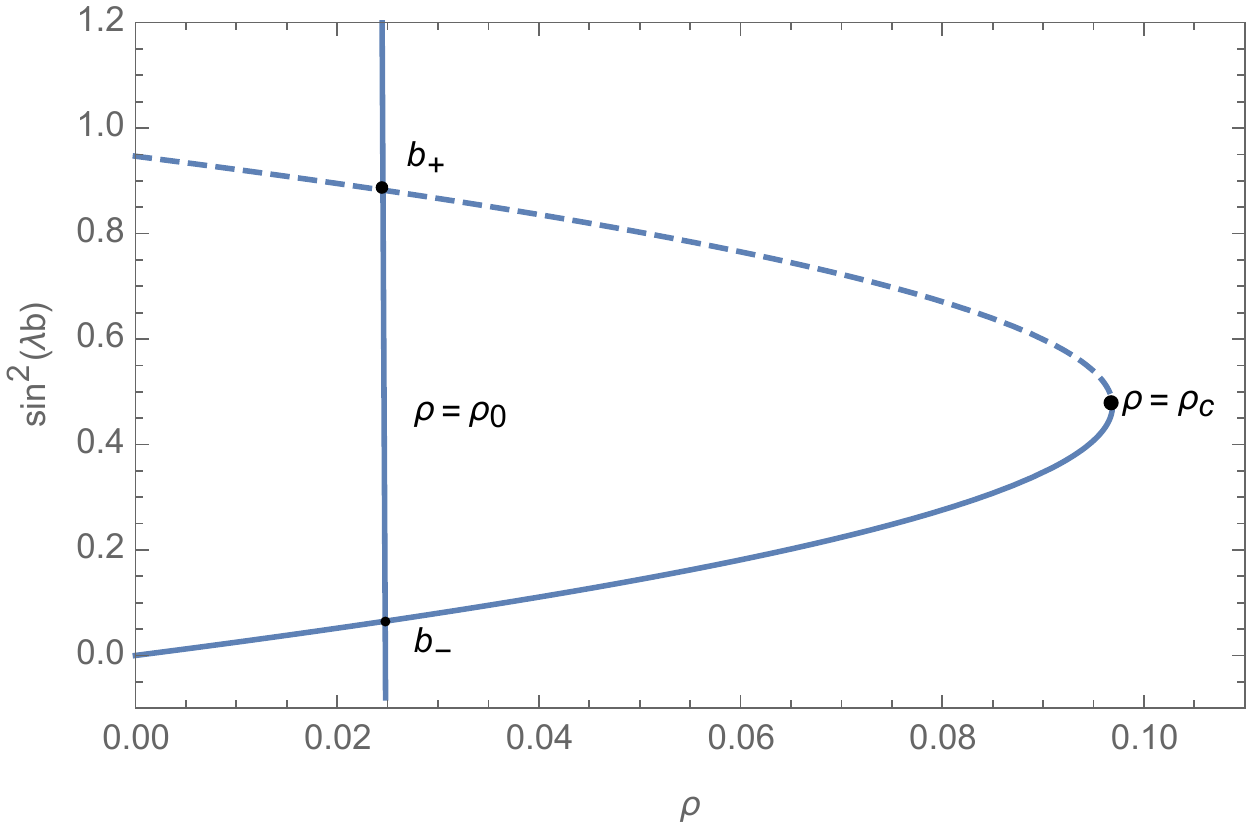}
}
\caption{In this figure, we plot Eq. (\ref{Hcd}). The top dashed line represents the $b_+$ branch while the bottom solid curve the $b_-$ branch. The two branches converge at the 
critical density $\rho=\rho_c$. For any choice of the  initial density $\rho=\rho_0 < \rho_c$,  which is represented by the solid vertical line in the graph, there are two solutions of 
$\sin^2(\lambda b)$ marked by  $b_+$ and $b_-$, respectively. In the actual simulations, $\rho_0$ will be chosen  in the range of $10^{-4}$ -- $10^{-6}$ in terms of the Planck units.}
\label{fig1}
\end{figure}

On the other hand, from the Hamiltonian constraint (\ref{HCD}), we can directly find the expression for the energy density $\rho\equiv \mathcal{H}_M/v$, 
given by
\bq
\lb{rho}
\rho=\frac{3}{8\pi G\lambda^2}\left(-\sin^2(\lambda b)+\frac{(\gamma^2+1)\sin^2(2\lambda b)}{4\gamma^2}\right),  
\eq
or inversely, 
\bqn
\lb{Hcd}
\sin^2(\lambda b_{\pm})= \frac{1\pm\sqrt{1-\rho/\rho_c}}{2(\gamma^2+1)},  
\eqn
where  $\rho_c \equiv3/[32\pi \lambda^2\gamma^2(\gamma^2+1)G]$. Note that,  in comparing with the maximum density  in LQC \cite{aps3,slqc}, the maximum density above  
is suppressed by a factor of $1/[4(\gamma^2+1)]$.  Then, for any given $\rho$, there are two sets of $b$, given by
\bqn
\lb{Hcd0}
 \lambda b_{\pm}&=& \epsilon_1 \sin^{-1}\left[\left(\frac{1\pm\sqrt{1-\rho/\rho_c}}{2(\gamma^2+1)}\right)^{1/2}\right] \nb\\
 && +  \epsilon_2 n \pi,  
\eqn
where $n = 0,  1, 2, ...$, and $\epsilon_{1, 2} = \pm 1$. Note that the choices of the signs of $\epsilon_{1, 2} $ are independent of the choice of the $b_{+}$ or $b_{-}$ branch. To emphasize this point,
which is important to our   later discussions to be carried out in the next two sections, 
here we had introduced $\epsilon_{1, 2}$.  
  At $\rho = \rho_c$, we have $\left.\sin^2(\lambda b_{-})\right|_{\rho_c} = \left. \sin^2(\lambda b_{+})\right|_{\rho_c}  = 1/[2(\gamma^2+1)]$, as shown in Fig. \ref{fig1}.
 
The function $\sin^2(\lambda b_{-})$ is a monotonically increasing function of $\rho$ [Cf. Fig. \ref{fig1}], and
\bqn
\lb{Hcd1}
\sin^2(\lambda b_{-}) 
=\begin{cases}
0, & \rho = 0,\\
\frac{1}{2\left(\gamma^2+1\right)}, &  \rho = \rho_c.
\end{cases}
\eqn
In contrast, $\sin^2(\lambda b_{+})$ is a monotonically decreasing function of $\rho$, and
\bqn
\lb{Hcd2}
\sin^2(\lambda b_{+}) 
=\begin{cases}
\frac{1}{\left(\gamma^2+1\right)}, & \rho = 0,\\
\frac{1}{2\left(\gamma^2+1\right)}, &  \rho = \rho_c.
\end{cases}
\eqn

It is interesting to note that in the standard LQC there exists only one branch of solutions of the Hamiltonian constraint (see Sec. IIIA), given by \cite{aps3}
\bq
\lb{LQCa}
\sin^2(\lambda b) = \frac{\rho}{\tilde \rho_c},
\eq
as the Lorentzian term, represented by the second one on the right-hand side of
Eq. (\ref{Ha}) is not explicit. Here $\tilde \rho_c$ denotes the maximum energy density in LQC: $\tilde\rho_c \equiv 3/[8\pi G \lambda^2\gamma^2] \simeq 0.41 \rho_{Pl}$. As a result, the evolution of the universe in LQC is symmetric with respect to the bounce at which we have
$\rho = \tilde\rho_c$ \cite{aps3}. As ${\rho}/{\tilde\rho_c} \ll 1$, Eq.(\ref{LQCa}) is identical to the $b_{-}$ branch with $\tilde\rho_c =4(\gamma^2+1)\rho_c$. As we show below, a fundamental difference between LQC and
the LQG cosmology studied in this paper is that in the latter the evolution of the universe is asymmetric with respect to the bounce. In fact, after taking the current cosmological constraints into account, the evolution must be starting
from the $b_{+}$ branch in the pre bounce phase until the bounce,  and afterward the evolution will be switched to the one described by   the $b_-$ branch [Cf. Fig. \ref{fig1}]. 

It should be also noted that in the LQG cosmology  the two branches of solutions which are found, share similar features studied earlier for higher-order modifications than the usual quadratic corrections to the Friedmann equation in  LQC \cite{singh-soni}.  
Our analysis in the following would actually yield a similar modified Friedmann equation and the multiple branches of the kind studied earlier in the above reference.

In addition,  Eq. (\ref{1.1}) is the generalized Friedmann equation, which, together with Eq. (\ref{1.1b}),  is equivalent to the Hamilton's equations (\ref{eomA}) and  (\ref{eomB}). To write the right-hand sides of
Eqs. (\ref{1.1}) and (\ref{1.1b}) in terms of $\rho$ and $P$ only, we need to distinguish the two branches $ b = b_{-}$ and $b = b_{+}$, when we try to solve the Hamiltonian constraint 
(\ref{HCD}) to obtain the variable $b$ in terms of $\rho$, as shown explicitly by Eq. (\ref{Hcd}).

\subsection{$b=b_{-}$}

Considering the branch with $b = b_{-}$, we find that
\bq
\lb{1.3}
\dot b= -4\pi G  \gamma (\rho+P),
\eq
and Eqs. (\ref{1.1}) and (\ref{1.1b}) take the forms
\begin{widetext}
\bqn
\lb{1.4a}
H^2 &=&\frac{8\pi G \rho}{3}\left(1-\frac{\rho}{\rho_c}\right)\Bigg[1  +\frac{\gamma^2}{\gamma^2+1}\left(\frac{\sqrt{\rho/\rho_c}}{1 +\sqrt{1-\rho/\rho_c}}\right)^2\Bigg], \\
\lb{1.4b}
\frac{\ddot a}{a} &=&-\frac{4\pi G}{3}\left(\rho + 3P\right)
  + \frac{4\pi G \rho}{3}\left[\frac{\left(7\gamma^2+ 8\right) -4\rho/\rho_c+\left(5\gamma^2 +8\right)\sqrt{1-\rho/\rho_c}}{(\gamma^2 +1)\left(1+\sqrt{1-\rho/\rho_c}\right)^2}\right]\frac{\rho}{\rho_c}\nb\\
  &&  + 4\pi G P \left[\frac{3\gamma^2+2+2\sqrt{1-\rho/\rho_c}}{(\gamma^2+1)\left(1+\sqrt{1-\rho/\rho_c}\right)}\right]\frac{\rho}{\rho_c}.
\eqn
\end{widetext}
From these two equations, it can be shown that  the energy conservation law still holds 
\bq
\lb{ecl}
\dot \rho+3H(\rho +P)=0.
\eq
In fact, along with the modified Friedmann equation (\ref{1.4a}) and the modified Raychaudhuri equation (\ref{1.4b}), the above conservation law forms a consistent set. It can be easily checked that as an alternative 
to the above derivation which confirms that the matter-energy conservation law is unmodified, 
we can obtain any of the modified FR equations 
by starting from the conservation law and using the modified Raychaudhuri or the modified Friedmann equation, respectively. 

When $\rho = \rho_c$, Eqs. (\ref{1.4a}) and (\ref{1.4b}) show that a quantum bounce occurs.  As a result,  the big bang singularity is resolved even in the framework of the full LQG, similar to LQC \cite{aps,aps3,slqc}. The bounce is accompanied by a phase of super-inflation which can be determined from  $\dot H$ using Eqs.(\ref{1.4a})-(\ref{1.4b}). These give:
\bqn
\lb{dotH1}
\dot H&=&\frac{4 G \pi (P+\rho)}{(1+\gamma^2)}\left(2\gamma^2+2\frac{\rho}{\rho_c}-3\gamma^2\sqrt{1-\frac{\rho}{\rho_c}}-1\right).\nb\\
\eqn
Therefore, regardless of the matter content, the equation $\dot H=0$ is always true at 
\bq
\rho=\frac{\rho_c}{8}\left(4-8\gamma^2-9\gamma^4+3\gamma^2\sqrt{8+16\gamma^2+9\gamma^4}\right) .
\eq
Denoting the energy density at which super-inflation starts as $\rho_s$, we find that for the $b_-$ branch $\rho_s=0.503228 \rho_c$ for $\gamma=0.2735$. In contrast, in LQC one obtains $\rho_s = \tilde\rho_c/2$ irrespective of the value of the Barbero-Immirzi parameter \cite{ps06}.

In the classical limit $\rho/\rho_c \ll 1$, Eqs. (\ref{1.4a}) and (\ref{1.4b}) reduce, respectively, to
\bqn
\lb{1.5a}
H^2 &\approx& \frac{8\pi G}{3}\rho, \\
\lb{1.5b}
\frac{\ddot a}{a}&\approx&-\frac{4\pi G }{3}\left(\rho+3P\right),  
\eqn
whereby  the standard relativistic  cosmology is recovered. 

In the following, we shall refer Eqs. (\ref{1.4a}) and (\ref{1.4b}) as the FR equations for the $b_-$ branch.  
Clearly,  the phase space of the solutions of the FR equations (\ref{1.5a}) and (\ref{1.5b}) only covers a part of the phase space of the solutions of the  
Hamilton's equations (\ref{eomA})-(\ref{eomB}). The other part of the phase space of the  solutions of  the Hamilton's equations
is covered by the FR equations obtained for the branch $b = b_{+}$.

\subsection{$b=b_{+}$}

In this case, Eq. (\ref{1.3}) still holds, while Eqs. (\ref{1.1}) and (\ref{1.1b}) take the following forms:
\begin{widetext}
\bqn
\lb{1.6a}
H^2 &=&\frac{8\pi G\alpha  \rho_\Lambda}{3}\left(1-\frac{\rho}{\rho_c}\right)\left[1+\left(\frac{1-2\gamma^2+\sqrt{1-\rho/\rho_c}}{4\gamma^2\left(1+\sqrt{1-\rho/\rho_c}\right)}\right)\frac{\rho}{\rho_c}\right], \\
\lb{1.6b}
\frac{\ddot a}{a} &=&- \frac{4\pi \alpha G}{3}\left(\rho + 3P - 2\rho_\Lambda \right)  +4\pi G\alpha P\left(\frac{2-3\gamma^2 +2\sqrt{1-\rho/\rho_c}}{(1-5\gamma^2)\left(1+\sqrt{1-\rho/\rho_c}\right)}\right)\frac{\rho}{\rho_c}\nb\\
&& - \frac{4\pi G\alpha \rho}{3}\left[\frac{2\gamma^2+5\gamma^2\left(1+\sqrt{1-\rho/\rho_c}\right)-4\left(1+\sqrt{1- \rho/\rho_c}\right)^2}{(1-5\gamma^2)\left(1+\sqrt{1-\rho/\rho_c}\right)^2}\right]\frac{\rho}{\rho_c},
\eqn
\end{widetext}
where $\alpha\equiv ({1-5\gamma^2})/({\gamma^2+1})$ and $\rho_\Lambda \equiv{3}/[{8\pi G\alpha \lambda^2(1+\gamma^2)^2}]$. 
It can be also shown that the conservation law (\ref{ecl}) still holds in the current case, forming a consistent set with Eqs. (\ref{1.6a}) and (\ref{1.6b}).  In addition, a quantum bounce happens at $\rho = \rho_c$, too. 

Let us now consider the $\dot H$ equation for this branch. Using Eqs. (\ref{1.6a})--(\ref{1.6b}), we get 
\bqn
\lb{dotH2}
\dot H&=&\frac{4 G \pi (P+\rho)}{(1+\gamma^2)}\left(2\gamma^2+2\frac{\rho}{\rho_c}+3\gamma^2\sqrt{1-\frac{\rho}{\rho_c}}-1\right),\nb\\
\eqn
from which one can find the root of $\dot H =0$, given by
\bq
\rho=\frac{\rho_c}{8}\left(4-8\gamma^2-9\gamma^4-3\gamma^2\sqrt{8+16\gamma^2+9\gamma^4}\right) ~.
\eq
For $\gamma=0.2735$, the energy density at which super-inflation starts is $\rho_s=0.376801 \rho_c$. In comparison to the $b_-$ branch, we find that the super-inflationary regime in LQG cosmology is also asymmetric with respect to the bounce. More precisely,  it is a bit longer in the pre-bounce phase.

A notable feature of this branch is that quantum geometry effects lead to an emergent cosmological constant \cite{DL17,adlp} with energy density $\rho_\Lambda$, and a modified Newton's coupling constant $G_\alpha := \alpha G$. It is interesting to note that a similar emergent cosmological constant arises in the loop quantization of the Kantowski-Sachs spacetime which is isometric to the Schwarzschild interior
 \cite{djs}. Let us note that if one considers just the modified Friedmann equation (\ref{1.6a}), then one may be tempted to absorb $\alpha$ in the $\rho_\Lambda$ to rescale the energy density of the emergent cosmological constant, and one may conclude that there is no modification to Newton's coupling constant. 
However, when one considers the modified Raychauhdri equation (\ref{1.6b}) then one finds that such an identification is not possible. Thus, quantum geometry leads to two independent effects in this branch: an emergent cosmological constant and a modified Newton's constant. It is therefore pertinent to consider the observational constraints separately for $\rho_\Lambda$ and $G_\alpha$ to understand the viability of this branch.

Before we come to the observational constraints on this branch, let us find the equations in the regime where $\rho \ll \rho_c$. Then, Eqs. (\ref{1.6a}) and (\ref{1.6b}) reduce, respectively, to
\bqn
\lb{1.7a}
H^2&\approx&\frac{8\pi G_\alpha}{3}\left(\rho+\rho_\Lambda\right), \\
\lb{1.7b}
\frac{\ddot a}{a}&\approx&-\frac{4\pi G_\alpha}{3}\left(\rho+3P - 2\rho_\Lambda\right).
\eqn
These take the form of the classical GR equations for ordinary matter and an emergent  positive cosmological constant but with a modified Newton's constant. We see that even in the above limit the quantum geometry effects are present in $G_\alpha$ and $\rho_\Lambda$. If this branch lies in the post-bounce universe where we live, then there are many obvious phenomenological  problems, including those arising  from $\rho_\Lambda$ and $G_\alpha$ on their own. In the following we point out some of these problems. First let us note that the energy density of the emergent cosmological constant is of the order of the Planck density, and in particular for $\gamma \approx 0.2375$, $\rho_\Lambda \approx 0.03 \rho_{\mathrm{Planck}}$. The value is of the same order as  the one deduced conventionally in quantum field theoretic arguments  for the vacuum energy in our universe. And as in the latter case, there is a problem to fit with the current observational constraints for the value of the cosmological constant/dark energy if we consider this branch as describing our expanding universe. The ``cosmological constant problem'' which we find in this branch yields a background spacetime that has a Planckian curvature and a Hubble length which is just over $2 \ell_{\mathrm{Planck}}$. Obviously, unless there is a way to reduce the value of this emergent cosmological constant, this background spacetime has a wide range of serious problems to permit a viable physical universe as we live in. One may wonder if it is possible to somehow tune the value of the emergent cosmological constant by allowing some parameters of quantum gravity to vary. Though this is not permitted in the present context, because apart from fundamental constants the only parameter $\gamma$ is also determined by black hole thermodynamics in LQG as $\gamma \approx 0.2375$, even if we assume such a possibility we quickly see that such a speculation does not work. For this let us note that keeping the fundamental constants unchanged, the only way to achieve this is by treating $\gamma$ as a free parameter and allowing it to take a large value. (Note that the area gap $\lambda^2$ is also determined in terms of $G$, $\hbar$ and $\gamma$). One sees that even before $\gamma$ barely doubles in value from the value determined by LQG, $\rho_\Lambda$ and $G_\alpha$ change sign to become negative. This sign change,  which occurs for $\gamma   >  (0.2)^{1/2}$, results in the set of the modified FR equations which in the limit $\rho \ll \rho_c$ can be interpreted as corresponding to Friedmann dynamics with 
a positive Newton's constant, a positive cosmological constant, but with pressure and matter energy density  becoming negative. Even if we ignore the inherent problems with the latter, let us further note that for large values of $\gamma$, the energy density $\rho_c$ also decreases. If one wishes to tune the emergent cosmological constant to an extremely small value consistent with current observations in the above setting, one pays the price of pushing the effective quantum gravity scale so many orders of magnitude below the Planck curvature scale that the model can be easily ruled out by various observations. 

If one takes a viewpoint as in cosmology that one can address the above ``cosmological constant problem'' at a later stage or one assumes that for some yet to be known reasons or in a more realistic model it may get solved, then one can still find that this branch is not favorable to describe our expanding universe. For this argument which needs to be independent of $\rho_\Lambda$, let us focus just on the change in Newton's constant due to the quantum geometry effects and assume there is no other change because of a large value of $\rho_{\Lambda}$ to the dynamical history of our universe. Let us recall that a change in the effective value of Newton's constant modifies the expansion rate and is equivalent to modifying the radiation components during the epoch of the big bang nucleosynthesis. In our case,  
for $\gamma \approx 0.2375$ as one takes in LQG, the change is $G_\alpha \simeq 0.68 G$. The Hubble rate decreases due to decrease in effective Newton's coupling constant when compared to the classical GR case.  As argued in  \cite{CL04}, such a decrease 
in the expansion rate results in a lower freeze-out temperature for weak interactions. 
This results in a lesser production of the primordial Helium, affecting the Helium to Hydrogen mass ratio. For a change in Newton's coupling constant the change in this ratio $\Delta Y_p$ becomes \cite{CL04}:
\begin{equation}
\Delta Y_p = 0.08 \left(\frac{G_\alpha}{G} - 1 \right).
\end{equation}
On the other hand,  the current margin of errors on the estimate of the mass ratio is $|\Delta Y_P| < 0.01$ \cite{COb}. Thus the resulting constraint is:
\bq
\lb{CD4}
\left|\frac{G_{{\alpha}}}{G} - 1\right| \le \frac{1}{8}.
 \eq
For $\gamma \approx 0.2375$ \cite{Mei04}, we find that $\left|{G_\alpha}/{G} -1 \right| \simeq 0.32 > 1/8$. As before, if we treat $\gamma$ as a parameter which can be varied then one finds that one needs $\gamma \le 1/\sqrt{47}$ for this branch to be not ruled out  by the current observations of the primordial Helium abundance.   Note that if one assumes $\gamma$ can be varied to take such a smaller value, one gets a larger value of $\rho_\Lambda$ making the earlier discussed problem worse. 
 
 Thus, we conclude from these separate arguments related to $\rho_\Lambda$ and $G_\alpha$ that the $b_+$ branch is unsuitable to describe an expanding universe such as ours if we use the observational constraints from either the cosmological constant or  Newton's constant. 
 However, it must be noted  that this branch of solutions can still exist in the pre-bounce epoch  ($t \le t_B$). Then, the open question is whether such a pre-bounce branch can serve as a viable  initial phase for a post-bounce universe described by the $b_-$ branch.

\section{Numerical Simulations of the Hamilton's Equations}
\label{Section3}
\renewcommand{\theequation}{3.\arabic{equation}}\setcounter{equation}{0}

In this section, we will concentrate ourselves on numerically solving   the Hamilton equations (\ref{eomA}) and  (\ref{eomB}) with the  Hamiltonian constraint (\ref{HCD}). 
To this purpose, let  us first consider a massless scalar field so that the matter part of the Hamiltonian is given by
\bq
\lb{3.1}
\mathcal {H}_M=\frac{\pi^2_\phi}{2v},
\eq
from which we find that 
\bqn
\lb{3.2a}
\dot \pi_{\phi} &=& \big\{\pi_{\phi}, \mathcal {H}_M\big\} =  0,\\
\lb{3.2b}
\dot \phi&=& \big\{\phi, \mathcal {H}_M\big\} =  \frac{\pi_\phi}{v},
\eqn
which show that  $\pi_\phi$ is a constant of motion. So, the 4-dimensional phase space now reduces to a 3-dimensional 
hypersurface $\pi_\phi = $ constant, say, $\pi_\phi^0$. Then,  Eqs. (\ref{eomA}) and (\ref{eomB}) become
\bqn
\lb{3.3a}
\dot v&=& \frac{3v\sin(2\lambda b)}{2\gamma \lambda}\Big\{(\gamma^2+1)\cos(2\lambda b)-\gamma^2\Big\}, \\
\lb{3.3b}
\dot b&=& \frac{3\sin^2(\lambda b)}{2\gamma \lambda^2}\Big\{\gamma^2\sin^2(\lambda b)-\cos^2(\lambda b)\Big\}\nb\\
&& -\frac{2\pi G\gamma {\pi_\phi^0}^2}{v^2},
\eqn
while the Hamiltonian constraint (\ref{HCD}) reduces to 
\bq
\lb{constraint}
\frac{3v}{8\pi G\lambda^2}\Bigg\{\sin^2(\lambda b)-\frac{(\gamma^2+1)\sin^2(2\lambda b)}{4\gamma^2}\Bigg\}+\frac{{\pi_\phi^0}^2}{2v} = 0.
\eq
It is interesting to note that  Eqs. (\ref{3.2b})-(\ref{3.3b}) are invariant under the rescaling, $v \rightarrow v/|\pi_\phi^0|$ for $\pi_\phi^0 \not= 0$. In addition, they also possess 
the  time-translation symmetry, $t  \rightarrow t + t_1$, where $t_1$ is a constant. This symmetry allows us to choose any value of $t$ as the initial time.

To solve Eqs. (\ref{3.2b})-(\ref{3.3b}), we need to specify the initial values for $\phi, \; b$ and $v$ at $t_0$.  However, because of the Hamiltonian constraint (\ref{constraint}), only one of the two ($b_0, v_0$) can be chosen freely, and the 
second one must satisfy this constraint. Here $b_0 \equiv b(t_0)$, $v_0 \equiv v(t_0)$,  etc. For example, if we choose to specify $v_0$,  then $b_0$ cannot be chosen freely any longer, as it must satisfy Eq.(\ref{constraint}). Then, in principle we can solve this equation to
obtain $b_0$. However,   such obtained $b_0$ is not unique, as it can be seen from Eq.(\ref{Hcd0}),  where for any chosen $v_0$ we have $\rho_0 = \left(\pi^0_{\phi}/\sqrt{2}v_0\right)^2$. 
To overcome this ambiguity, we can specify $b_0$ and then solve Eq. (\ref{constraint}) to obtain $v_0$,
\bq
\lb{v0}
v_0 =\left|\pi_\phi^0\right| \left(\frac{16\pi\lambda^2\gamma^2 G/3}{(\gamma^2+1)\sin^2(2\lambda b_0) - 4\gamma^2\sin^2(\lambda b_0)}\right)^{1/2}. 
\eq
Note that in writing down the above equation, we had chosen the positive  sign, as $v_0$ represents the initial volume. Once the  Hamiltonian constraint (\ref{constraint}) is satisfied initially, it will be satisfied at any other moment, as it is
conserved $\dot{\mathcal {C}} \approx 0$. 
Therefore, to solve the dynamical equations (\ref{3.2b}) - (\ref{3.3b}), we just need to specify the initial data $(\phi_0, b_0)$, from which $v_0$ is determined by
Eq. (\ref{v0}). Once $(\phi_0, b_0, v_0)$ is specified, the solution of the dynamical equations (\ref{3.2b}) - (\ref{3.3b}) will be uniquely determined. In particular, at the bounce, we have
\bqn
\lb{AtB}
\left. \dot \phi\right|_{\rho = \rho_c} &=&    \frac{\pi^0_\phi}{v_c}, \quad \left. \dot v\right|_{\rho = \rho_c}  = 0,\nb\\
\left. \dot b \right|_{\rho = \rho_c} &=& - \frac{3} {8 \lambda^2 \gamma(\gamma^2+1)} - \frac{2\pi G\gamma {\pi_\phi^0}^2}{v_c^2} < 0,
\eqn
where $v_c \equiv \left. v\right|_{\rho = \rho_c} > 0$. The above expressions show that $\dot{\phi}, \; \dot{v}$ and $\dot{b}$ are all well defined at the bounce, so the functions $\phi(t),\; v(t)$ and $b(t)$ must be well behaved across the
bounce.  In addition, it can be also shown that their second derivatives are also well defined at the bounce, so these functions must be at least $C^2$ across the bounce. 
It is in this sense we say that the evolution of the universe across the bounce is unique. 
 
To solve these dynamical equations numerically, we shall  set $G=1$ and $\ell_{\mathrm{Planck}} = 1$, from which we find that $\lambda = \left(4\sqrt{3}\; \pi \gamma\right)^{1/2} \simeq 2.2736$.  Therefore, in our numerical simulations 
both the energy density and time are in the Planck units. 

In addition, we would also like to compare our results with those obtained from LQC and GR. To this goal, we shall choose the same initial conditions in all these three theories. We briefly summarize the 
dynamical equations needed for LQC and GR in the following.

\subsection{LQC}

In the framework of LQC, the effective Hamiltonian for the spatially flat homogeneous and isotropic FLRW spacetime is given by \cite{aps3}
\bq
\lb{3.4a}
\mathcal{H}_{LQC}=-\frac{3v\sin^2(\lambda b)}{8\pi G \gamma^2\lambda^2}+\frac{\pi^2_\phi}{2v},
\eq
from which we obtain the following Hamilton's equations,
\bqn
\lb{3.4b}
\dot \phi&=&\frac{\pi_\phi}{v}, \quad \dot \pi_\phi=0, \\
\lb{3.4c}
\dot v&=&\frac{3v}{2\lambda \gamma}\sin(2\lambda b),  \\
\lb{3.4d}
\dot b&=&-\frac{3\sin^2(\lambda b)}{2 \gamma \lambda^2}-\frac{2 \pi G \gamma \pi^2_\phi}{v^2} ~.
\eqn
From Eq. (\ref{3.4b}) we can see that $ \pi_\phi = $ constant. Again, by rescaling $v$ we can always set $ \pi_\phi$ to unity.

Note that the bounce happens at $v = v_b$ when $\sin^2(\lambda b) = 1$, at which we have $\dot \phi=\pi^0_\phi/v_b$, 
$\dot v = 0$,  $ \dot b = -3/(2 \gamma \lambda^2)-2 \pi G \gamma \pi^2_\phi/v_b^2 < 0$, where $v_b > 0$. Similar to the LQG cosmology, now these functions and their derivatives up to the second-order are all well defined
and continuous across the bounce. As a result, the evolution of the universe is also unique across the bounce.  
 
\subsection{Classical GR}
In GR, the Hamiltonian for the spatially flat homogeneous and isotropic FLRW spacetime in terms of $b$ and $v$ variables is,
\bq
\mathcal{H}_{GR}=-\frac{3v b^2}{8\pi G\gamma^2}+\frac{\pi^2_\phi}{2v}, 
\eq
which results in the following Hamilton's equations:
\bqn
\lb{3.5a}
\dot \phi&=&\frac{\pi_\phi}{v}, \quad \dot \pi_\phi=0, \\
\lb{3.5b}
\dot v&=&\frac{3vb}{\gamma},  \\
\lb{3.5c}
\dot b&=&-\frac{3b^2}{2\gamma}-\frac{2 \pi G \gamma \pi^2_\phi}{v^2}, \\
\lb{3.5d}\mathcal{H}_{GR}&=&-\frac{3v b^2}{8\pi G\gamma^2}+\frac{\pi^2_\phi}{2v} =  0.
\eqn
The Hamiltonian equations in GR can also be obtained from Eqs. (\ref{eomA})-(\ref{HCD}) or from Eqs. (\ref{3.4a})-(\ref{3.4c}) by taking the classical limit.

Using the classical Hamilton's equations, we obtain
\bqn
v(t)&=&\sqrt{12\pi G}\; \pi_{\phi}^0\;   (t +t_1), \nb\\
b(t) &=&\frac{\gamma}{3(t+t_1)}, 
\eqn
where $\pi_{\phi}^0$ and $t_1$ are two integration  constants, and without loss of the generality, we can alway set $t_1$ to zero, so  the volume becomes zero at $t= 0$. Then,  
at $t = t_0$ we have 
\bq
\lb{3.6}
b_0 = \frac{\gamma}{3t_0}, \quad v_0=\sqrt{12\pi G}\; \pi_{\phi}  t_0. 
\eq
In addition, the energy density is given by
\bq
\lb{3.7}
\rho=\frac{3b^2}{8\pi G\gamma^2}.
\eq
Note that at $t= 0$ we have $b = \infty$, and the energy density becomes unbounded at this moment - the big bang singularity.

 \subsection{Numerical Simulations}

 Having discussed the Hamilton's equations for LQG cosmology, LQC and GR, in this subsection we will present some representative results on the numerical simulations in these three different theories.
Specifically,  our numerical simulations are chosen to start, without  loss of the generality, at the time $t_0=11.5$ (in Planck units) with
\bq
\lb{3.8}
\pi^0_\phi=1, \quad \quad \rho_0=10^{-4}, 
\eq
and then we integrate the Hamilton's equations   
backward to the pre-bounce region. 
Note that although this time and some relevant quantities are also denoted  with the subscript ``0", 
they do not represent the physical initial conditions of the system, instead its final state, as we are integrating backward in time.  However, the uniqueness theorem of the ordinary differential equations  guarantees that  the trajectory is uniquely determined 
by the initial condition $(b_0, v_0, \phi_0)$. Therefore, the time $t_0=11.5$ and the corresponding quantities, such as  $\pi^0_\phi, \; \rho_0, \; v_0$ and $b_0$,  actually represent the starting moment of our numerical simulations. It is exactly in this sense 
we refer to them as ``the initial conditions."

\begin{figure}
{
\includegraphics[width=8cm]{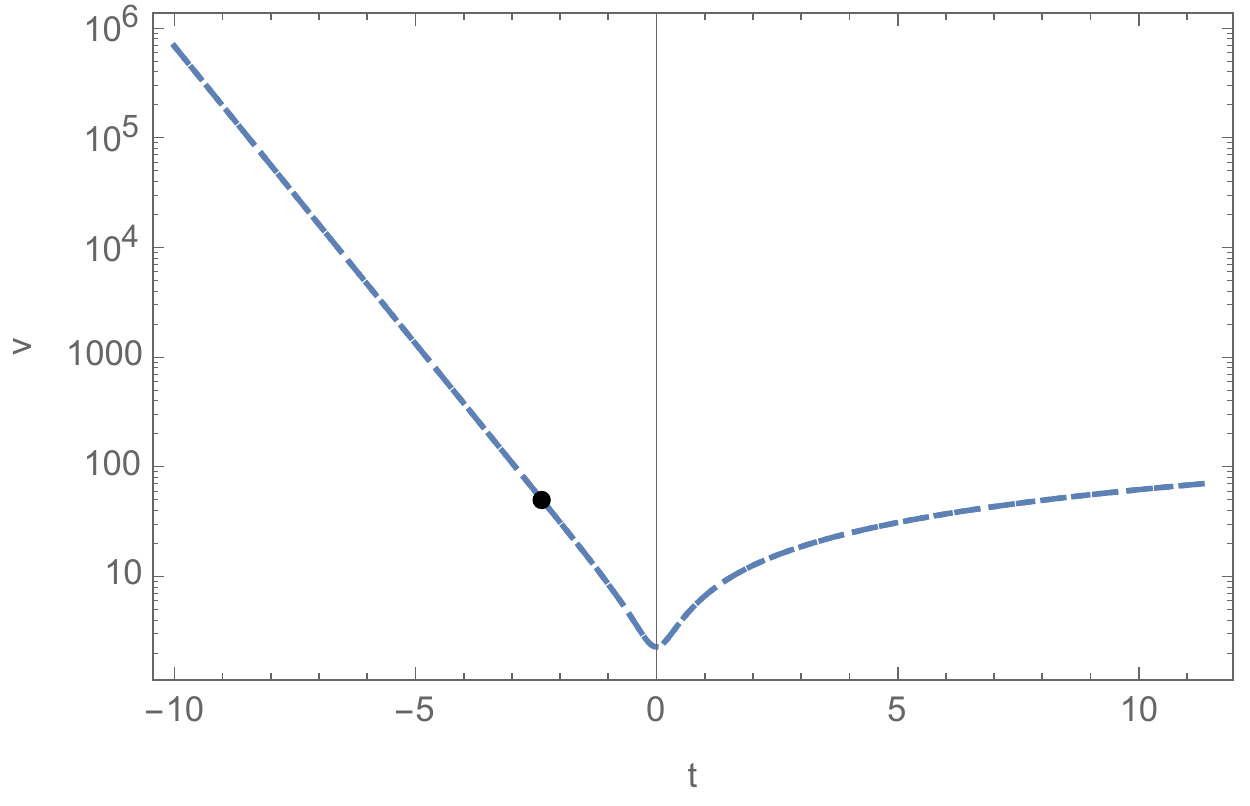}
\includegraphics[width=8cm]{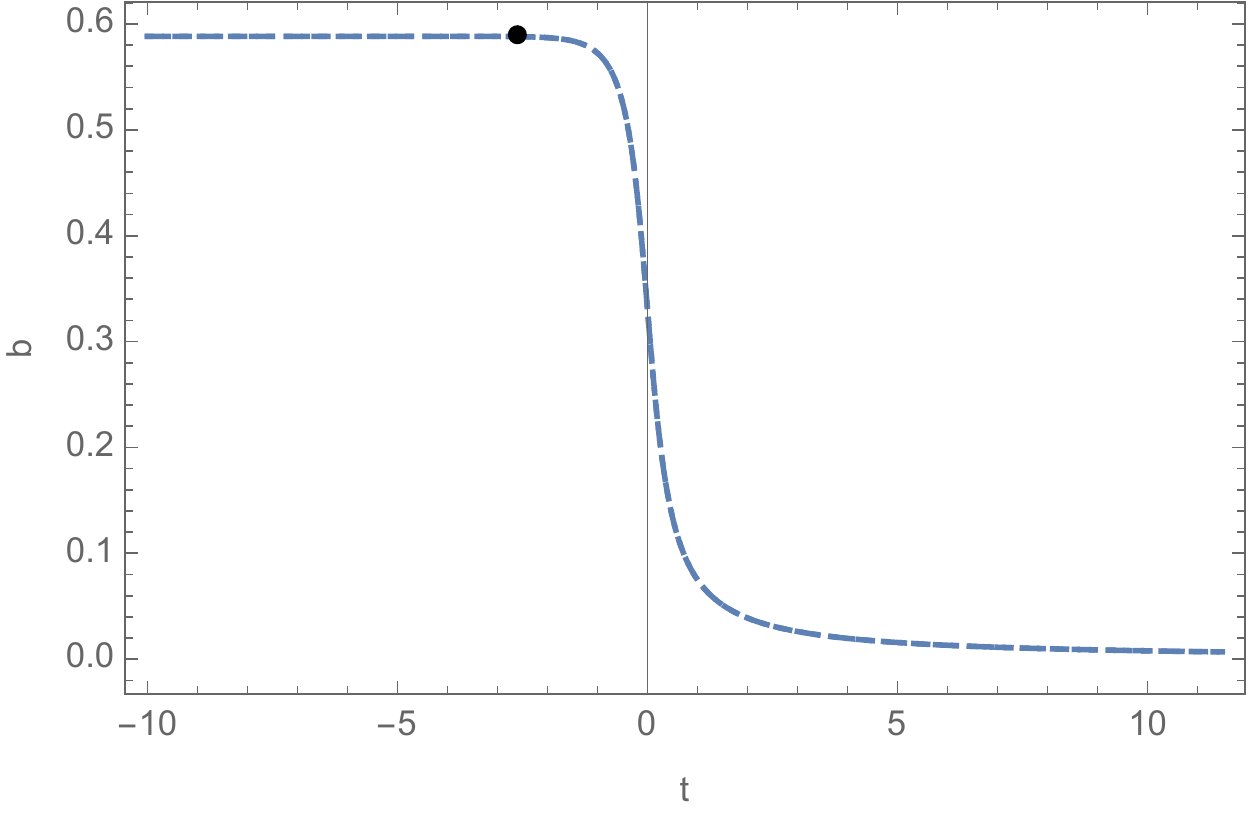}
\includegraphics[width=8cm]{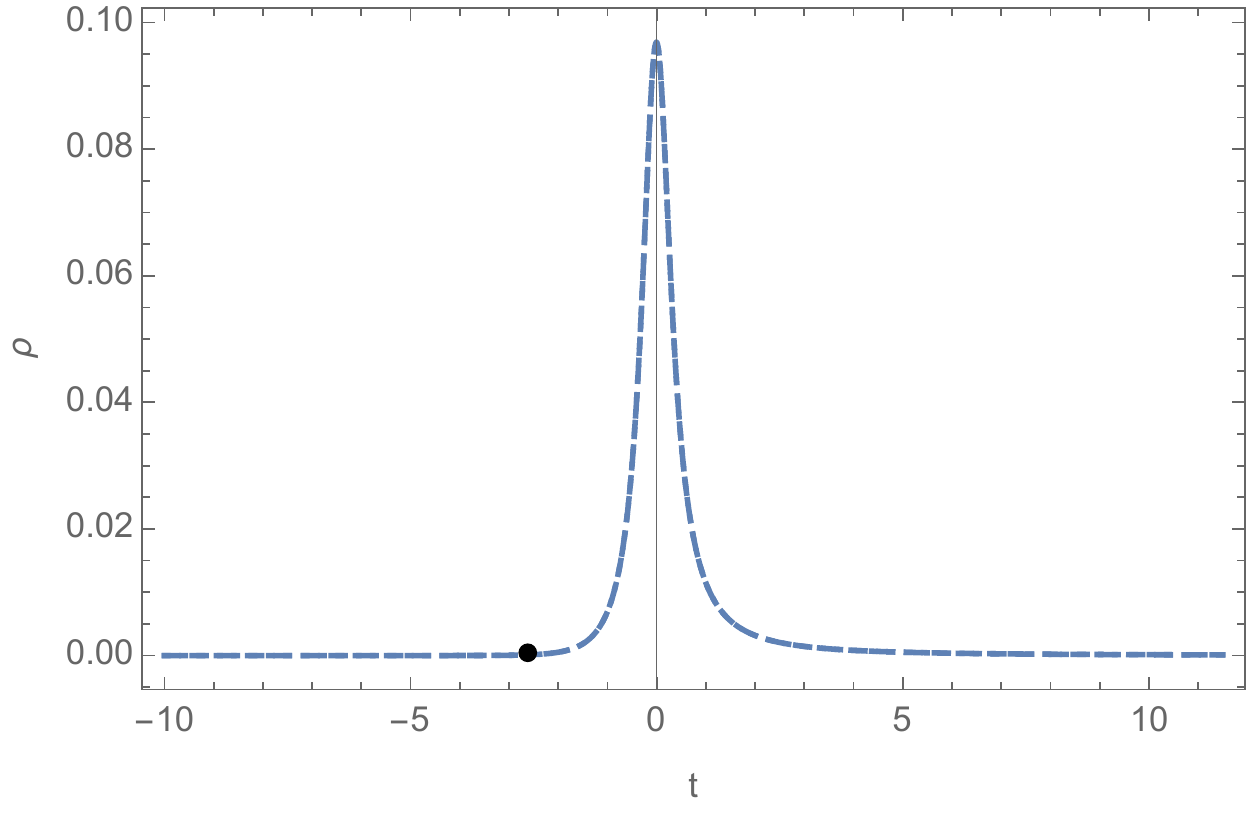}
}
\caption{In this figure, the initial time of our numerical simulations is chosen at 
$t_0=11.5$ and the final time is set to $t_f=-10$. Hence, the bounce occurs at $t_B=-0.004$,
and at $t_0=11.5$,   we have   $v_0\approx 70.71$ and  $b_0^-\approx 0.00688$. }
\label{fig2}
\end{figure}

Once ``the initial conditions"  given by Eq. (\ref{3.8}) are set, the initial volume $v_0$ is uniquely fixed by 
\bq
\lb{3.8a}
v_0=|\pi^0_\phi|\sqrt{\frac{1}{2\rho_0}},
\eq
which is identical to the value given in Eq. (\ref{v0}).  For $\pi^0_\phi$ and $\rho_0$ given by Eq.(\ref{3.8}), we have $v_0\approx 70.71$. 
 Then, from Eq.(\ref{Hcd0}) we find that
 \bq
 \lb{3.8aa}
 b^0_{\pm} = \epsilon_1 \left(\begin{array}{c}
 0.588\\
 0.00688
 \end{array}
 \right)   + \epsilon_2 \left(\frac{ n \pi}{\lambda}\right).
 \eq
Choosing $\epsilon_1 = +1$ and $ n = 0$, we find that      ${b}^0_-\approx 0.00688$ and ${b}^0_+\approx 0.588$.
 These two different values    correspond to two different times along the same trajectory, uniquely determined by
 initial conditions, as discussed above, and  illustrated  in  Fig. \ref{fig1}. To see this more clearly, let us first  note that  $b(t)$ is always decreasing for any matter that satisfies the weak energy condition, as can be seen from Eq.(\ref{1.3}). 
Thus, for the choice  $\epsilon_1 = +1,\; n = 0$, the evolution of the universe must be described by   the segment $B\rightarrow A \rightarrow O$ in Fig. \ref{fig6}, that is, the universe starts in the pre bounce phase, in which it is described by the
$b_+$ branch solutions. As the universe is contracting $\dot{v} < 0$, the energy density is increasing until the point $\rho = \rho_c$, at which we have $\dot{v} = 0$ and $\dot{b} < 0$, as shown previously. Right after the bounce, the universe will
follow the curve described by the $b_-$ branch solution as shown in Fig. \ref{fig1}, or the segment $B\rightarrow A \rightarrow O$ in Fig. \ref{fig6}, because $\dot{b}$ is always negative, as shown in Eq.(\ref{3.8aa}), and due to the smoothness of 
the functions $\phi, \; v$ and $b$ across the bounce, as discussed above. In particular, the evolution of the universe after the bounce cannot be described continuously by the $b_+$ branch solution, since if this were the case, then $\dot b$ would be
increasing as shown in  Fig. \ref{fig1}, which contradicts   Eq.(\ref{1.3}). 
Therefore, in the current case we must choose $\left(b^{-}_0, v_0, \phi_0\right)$ as ``the initial condition".  
   Recall that we are integrating the system backwards, and starting  the evolution in the post bounce phase ($t = t_{0} = 11.5 > 0$). 
This can be seen clearly from Figs. \ref{fig1} and \ref{fig6}, and is further illustrated  in Fig. \ref{fig2}, from which we can see that  the solutions determined by  $b_0^+$ and $b_0^-$ are indeed located in the same trajectory. 
The only difference is that the state prescribed by $b_0^+$ lies in the pre-bounce phase, the de Sitter phase, described by Eqs.(\ref{1.7a}) and (\ref{1.7b}), while the state corresponding to $b_0^-$ in the post-bounce phase. Along the trajectory,   we find $b$ increases (backwards in time) to the value $b_+$ 
at $t^0_+\approx -2.658$,  which is shown by the black dots in  Fig. \ref{fig2}.

\begin{figure}
{
\includegraphics[width=8cm]{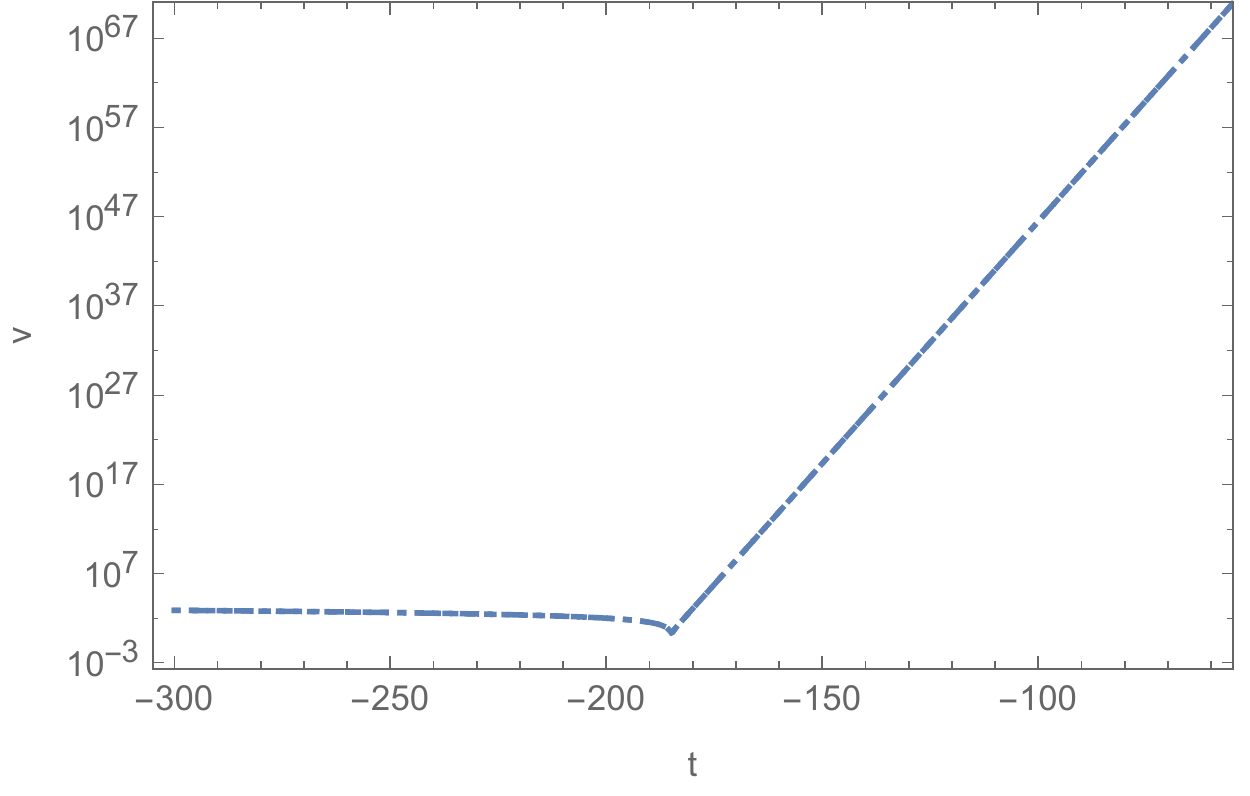}
\includegraphics[width=8cm]{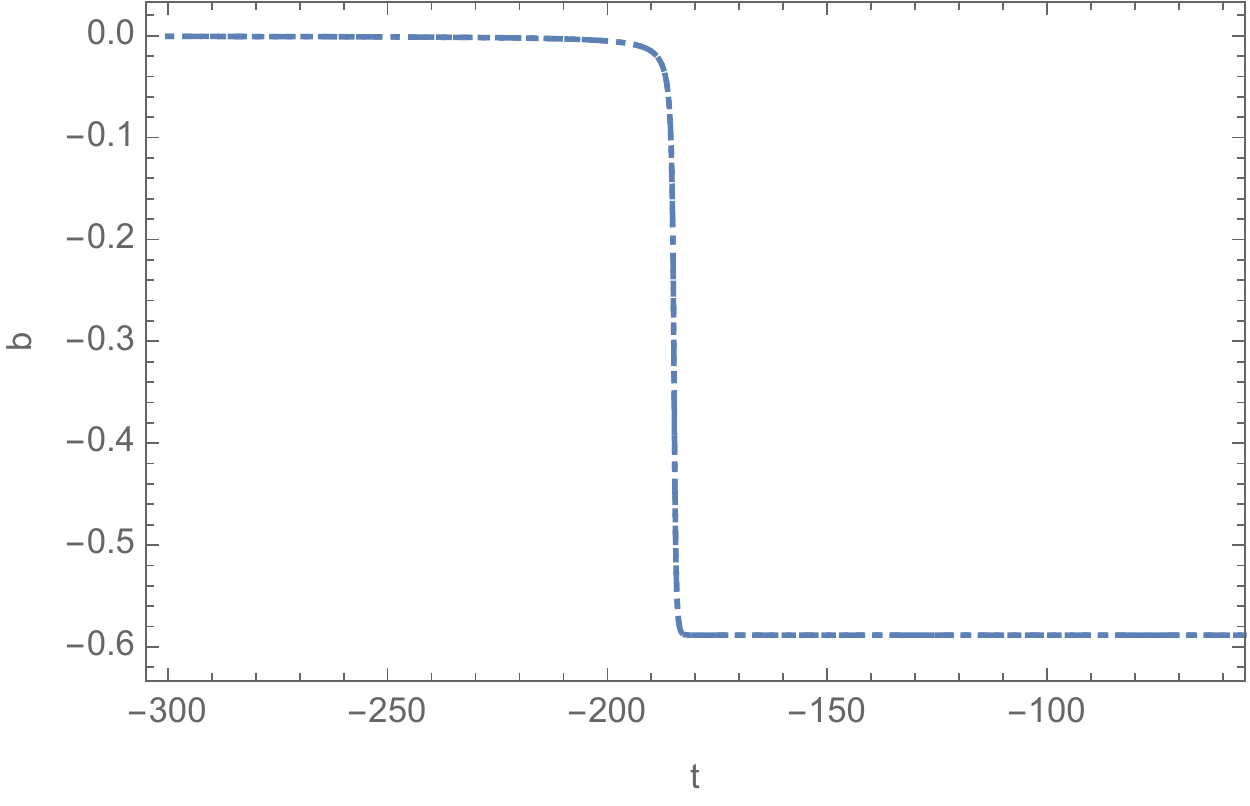}
\includegraphics[width=8cm]{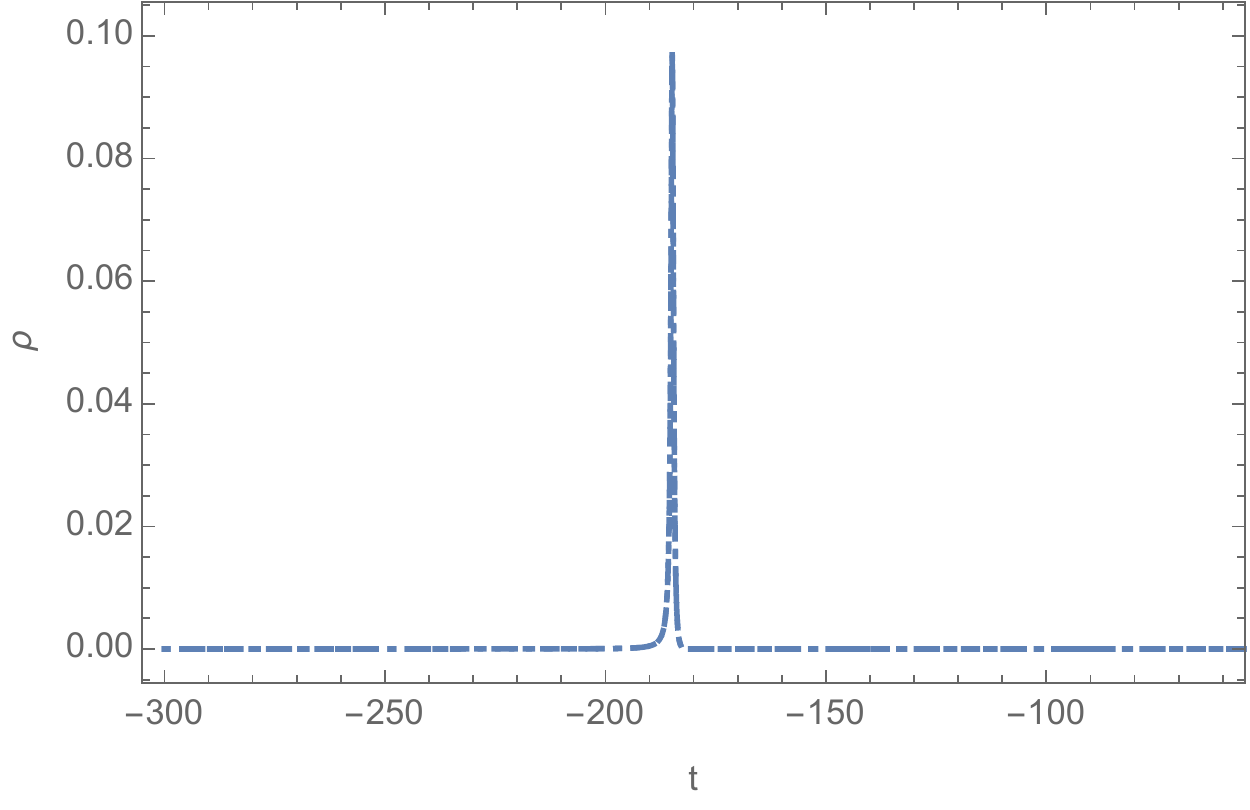}
}
\caption{In this figure, the initial time of our numerical simulations is chosen at 
$t_0=-300$ with $\rho_0=10^{-6}$ and $\pi^0_\phi=1$. Hence, the bounce occurs at $t_B=-185$, the de Sitter space is now located in the post bounce phase,
 the evolution of the universe in this figure is represented by the segment $O\rightarrow E\rightarrow F$ in Fig. \ref{fig6}. }
\label{fig}
\end{figure}

On the other hand, if we choose  $\epsilon_1 = -1,\; n = 0$, we find that   ${b}^0_-\approx - 0.00688$ and ${b}^0_+\approx - 0.588 < {b}^0_-$. Then, from Eq.(\ref{1.3}) we can see that now the pre-bounce phase must be described by the 
 $b_-$ branch solution. As the universe is contracting, the energy density increases until the point where $\rho = \rho_c$, at which we have $\dot{b} < 0$ and $\dot{v} = 0$, which is represented by the point $E$ in Fig. \ref{fig6}. Across the bounce,
 the universe will be described by the $b_+$ branch solution, and the whole process is described by the  segment $O \rightarrow E \rightarrow F$ in Fig. \ref{fig6} and  depicted in Fig. \ref{fig}. As pointed out previously, this possibility  is already ruled out by current  observations.

Similarly, if we choose $\epsilon_1=-1,\; \epsilon_2 = 1$ and $n=1$, the system will evolve along the segment $J\rightarrow D\rightarrow C$ in Fig. \ref{fig6},
 while if we choose $\epsilon_1=1,\; \epsilon_2 = -1$ and $n=1$, the system will evolve along the segment $G\rightarrow H\rightarrow I$.

\begin{figure}
{
\includegraphics[width=8cm]{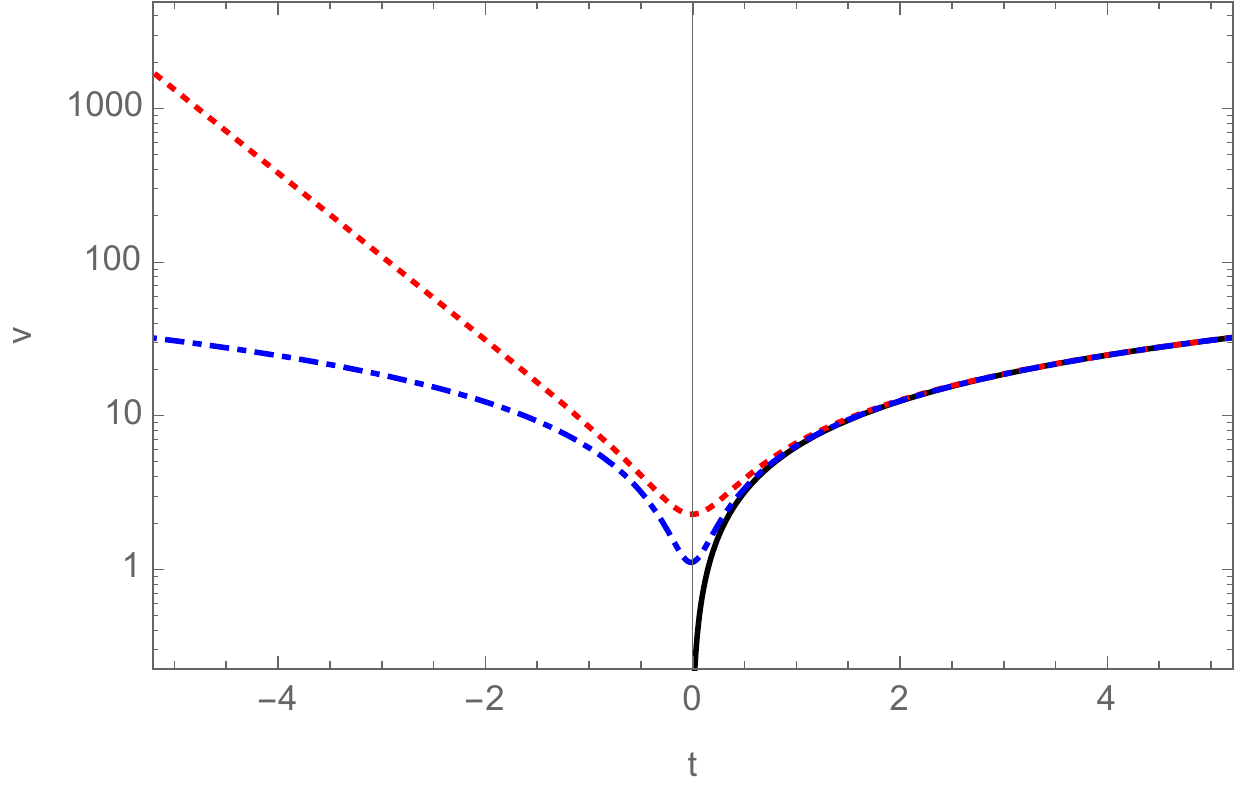}
\includegraphics[width=8cm]{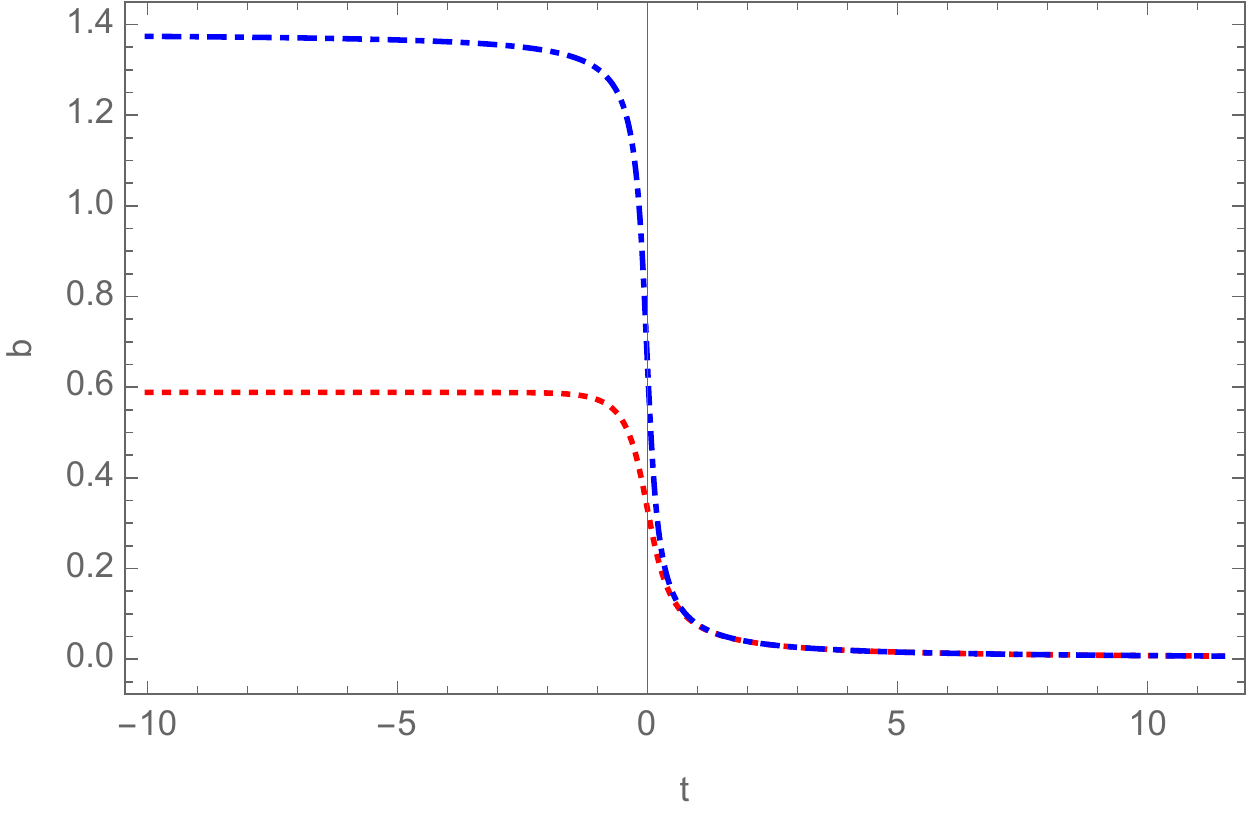}
\includegraphics[width=8cm]{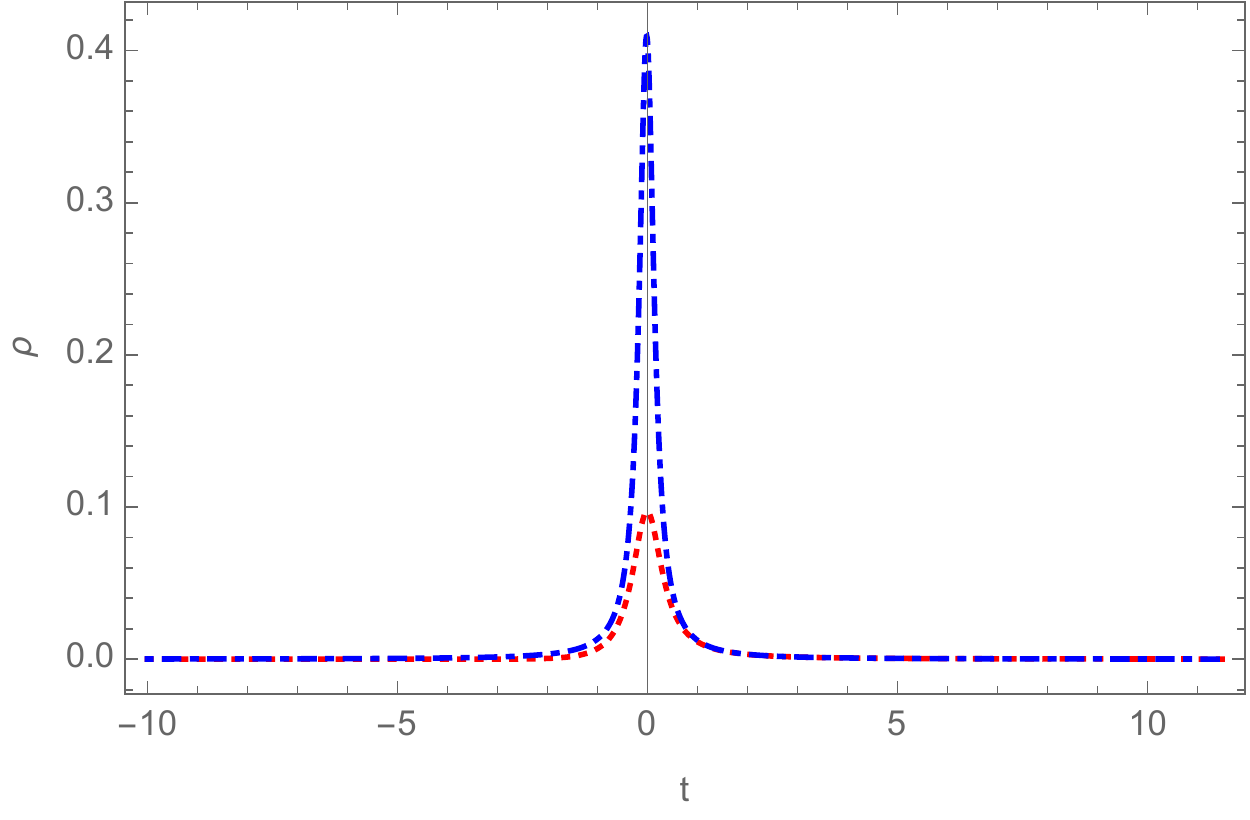}
}
\caption{In this figure, we compare the results from three different models. The black solid straight line is the result from the classical theory of GR, 
the red dotted line is from the full LQG cosmology and the blue  dot-dashed curve is from LQC. The initial conditions are chosen the same as in the Fig. \ref{fig2}.  }
\label{fig3}
\end{figure}

It is interesting to note that  the Hamilton's equations (\ref{3.3a})-(\ref{3.3b}) will generate similar results as those given by LQC  only in the phase where the evolution of the universe is described by the $b_-$ branch solutions.
In   Fig. \ref{fig3} we show the case where the universe in the LQG cosmology starts with the evolution described by the $b_-$ branch solutions in the post-bounce phase, along with the same initial conditions for LQC and GR. The GR solution ends at $v=0$ where the singularity is met. Note that there is another GR solution (not shown) for the $t<0$ which is disjoint with the shown solution. The bounce in LQC is symmetric. That is the post-bounce as well as the pre-bounce phase approaches classical GR solution for massless scalar field in the spatially flat FLRW spacetime asymptotically. Though, the LQC and LQG cosmology solutions both bounce and agree quite well in the post-bounce phase, there is striking disagreement between the two in the pre-bounce phase. For all of these simulations, we carefully monitored the validity of the Hamiltonian constraint, ${\cal{C}} \approx 0$, and the numerical errors were negligible. (This turns out to be true for all the figures presented in this paper.).

\begin{figure}
{
\includegraphics[width=8cm]{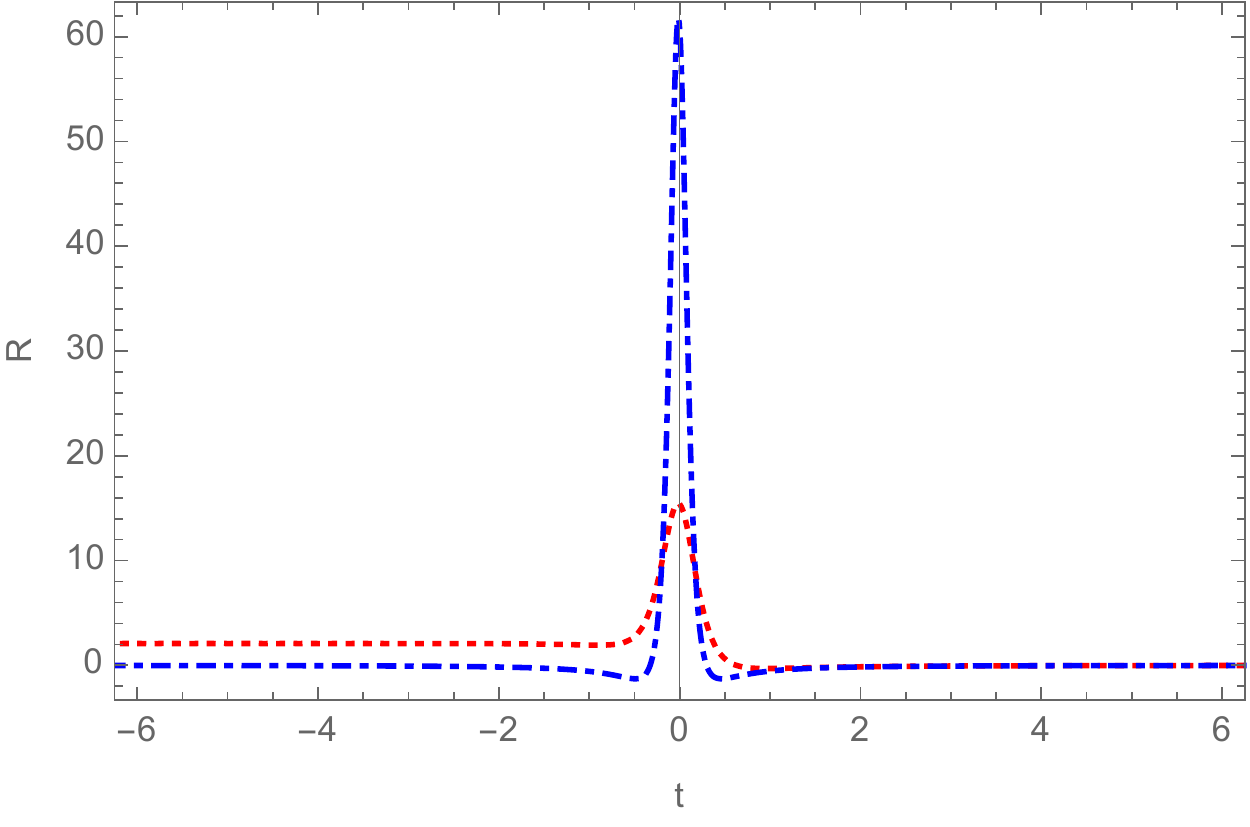}
\includegraphics[width=8cm]{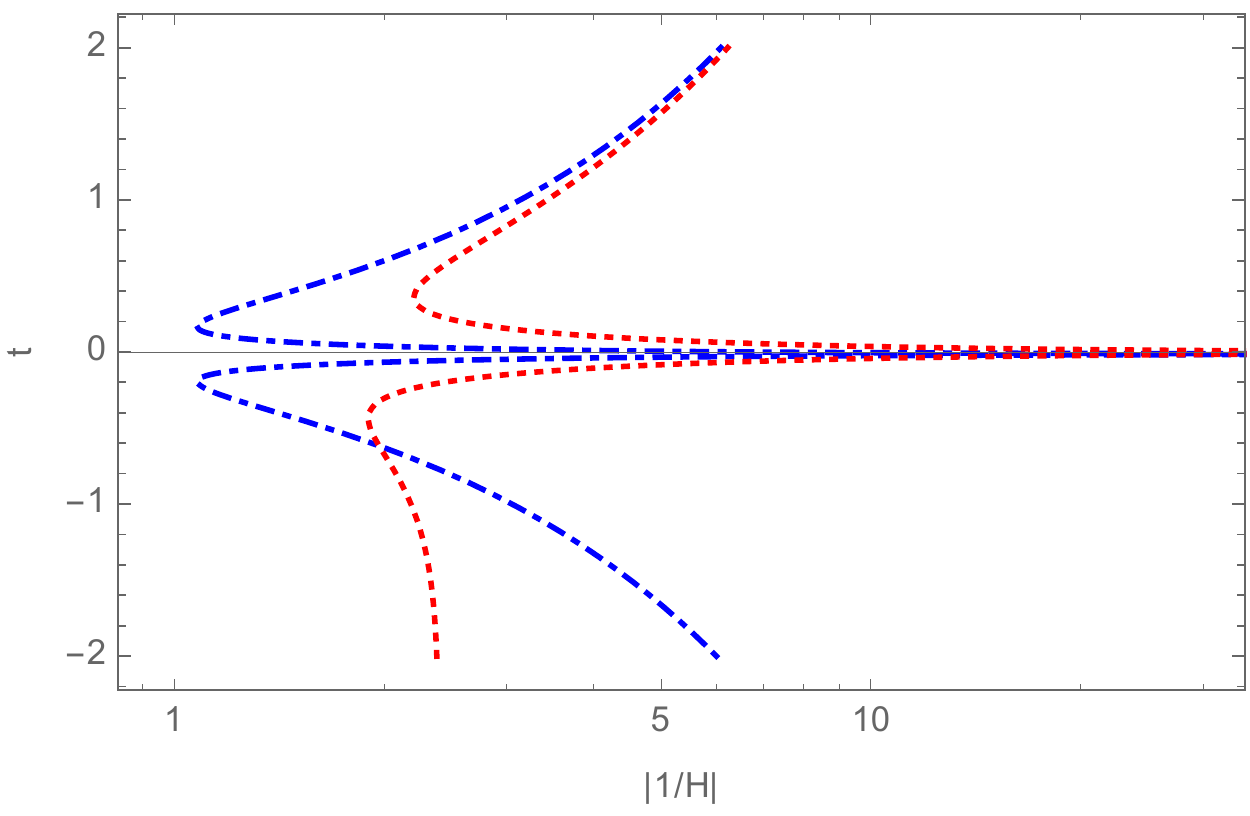}
}
\caption{With the same ``initial" conditions as adopted in Fig. \ref{fig3}, we show the behavior of the Ricci scalar and inverse Hubble rate near the bounce in Planck units. The LQG cosmology results are shown by 
the red dotted curve, and the LQC results  are depicted by the blue dot-dashed curve. }
\label{fig5}
\end{figure}

Finally, in Fig. \ref{fig5}, we   show the behavior of the Ricci scalar and the inverse Hubble rate near the bounce for both LQC and the LQG cosmology, with  the same ``initial" conditions as adopted in Fig. \ref{fig3}. From the curves of the Ricci scalar we can see that the curvature in LQG is much weaker than that in LQC at the bounce. In the pre-bounce branch, Ricci scalar takes a Planckian value at early times.  The Hubble length also has an interesting behavior. In the case of LQG cosmology, on one side of the bounce (for large negative times in the figure), the magnitude of the Hubble length takes Planckian value. On the post bounce side, its behavior is same as in LQC and it increases to larger values as the universe expands in the classical regime.

\section{Numerical Simulations of the FR equations}
\label{Section4}
\renewcommand{\theequation}{4.\arabic{equation}}\setcounter{equation}{0}

In Section II,  starting from the effective Hamiltonian, we  obtained dynamical equations in the form of the Hamilton's equations and the FR equations for the LQG cosmology. In the previous section, 
we studied the evolution of the systems using the Hamilton's equations, in which we paid particular attention on the consistency of Hamilton's equations (\ref{eomA}) and (\ref{eomB}) and the $b_{\pm}$ branches given by
Eq.(\ref{Hcd}). An ambiguity is introduced when we express this set of equations in terms of the expansion factor $a$, energy density $\rho$ and pressure $P$ of the fluid presented in the universe. It is remarkable 
that this ambiguity arises only in the LQG cosmology, and does not exist in either LQC or GR. Therefore,  care must be taken when analyzing dynamics using the FR equations for the LQG cosmology. 
In particular, we find that to be consistent with the Hamiltonian evolution, 
the FR equations have to switch the branch at the bounce from one to the other, as required by Eq.(\ref{1.3}), which shows that the function $b$ is always 
decreasing for any matter that satisfies the weak energy condition, as one can see from Fig. \ref{fig1}. In particular, at the bounce Eq.(\ref{1.3}) rules out the possibility that
the universe on both sides of the bounce is described by the same branch solutions, as in this the above equation will be violated either right after  or right before the 
bounce, as can be seen from Fig. \ref{fig1}.

In this section we continue such studies but by focusing  on the FR equations, which have
 two  sets of equations in the LQG cosmology, given, respectively, by Eqs. (\ref{1.4a})-(\ref{1.4b}), and  Eqs. (\ref{1.6a})-(\ref{1.6b}).  In the case of LQC, using Eqs.(\ref{3.4a})-(\ref{3.4d}), we can obtain the 
 the modified FR equations:
 \bqn
\lb{3.8ca}
H^2 &=& \frac{8\pi G}{3}\rho\left(1 - \frac{\rho}{\tilde\rho_c}\right),  \\
\lb{3.8cb}
\frac{\ddot a }{a}&=& -\frac{4\pi G}{3}\rho\left(1-\frac{4\rho}{\tilde\rho_c}\right)-4\pi G P\left(1-\frac{2\rho}{\tilde\rho_c}\right). ~~~~~~~~
\eqn
Here $\tilde\rho_c \equiv 3/[8\pi G \lambda^2\gamma^2] \simeq 0.41 \rho_{Pl}$. Using the above equations, we can easily obtain the $\dot H$ equation which is:
\bq
\dot H= -4 \pi G \left(P+\rho\right)\left(1-\frac{2\rho}{\tilde\rho_c}\right),
\eq 
from which we learn that, as long as the weak energy condition is satisfied, the super-inflationary phase starts at $\rho_s=\tilde\rho_c/2$ in the pre-bounce phase and ends when the energy density drops to $\rho_s$ again after the bounce. The super-inflationary regime is symmetric across the bounce and depends on the Barbero-Immrizi parameter only via $\tilde \rho_c$.  This is in contrast to the LQG cosmology where the super-inflationary phase is asymmetric and depends on the value of 
the Barbero-Immrizi parameter explicitly in addition to the dependence in bounce density $\rho_c$.

In contrast to the LQG cosmology, 
for  $ t \rightarrow \pm \infty$, the above equations lead to the same asymptotic behavior. One does not need two different sets of the FR equations to describe dynamics across the bounce in LQC. In comparing the above dynamical equations with 
Eqs.(\ref{1.4a})  and (\ref{1.4b}), we see that 
in the post-bounce, the differences in evolution in LQC and LQG cosmology become negligible as energy density decreases. However, in the pre-bounce phase their physics is strikingly different. In particular, in  this phase the LQG cosmology is described by Eqs.(\ref{1.6a})  and
(\ref{1.6b}), which asymptotically approach to a universe with a different coupling Newtonian constant $G_{\alpha} [= \alpha G]$ and an effective cosmological constant $\rho_{\Lambda}$, given by Eqs.(\ref{1.7a}) and (\ref{1.7b}).  In LQC, there is no emergent de Sitter behavior in the pre-bounce phase and as in the post-bounce phase one obtains a general relativistic spacetime when volume becomes large.

In the case of the classical GR, using   Eqs.(\ref{3.5a})-(\ref{3.5d}), we obtain the FR equations:
\bqn
\lb{4.1a}
&& H^2 = \frac{8\pi G}{3}\rho,\\
\lb{4.1b}
&&\ddot a = -\frac{4 \pi G}{3} \left(\rho + 3 P \right),
\eqn
which yield the conservation law Eq.(\ref{ecl}). For the massless scalar field we have $P = \rho$, for which the above equations have the solutions,
\bq
\lb{4.2} 
a = a_1\left[\pm(t+t_1)\right]^{1/6}, \quad \rho = \frac{\rho_1}{a^6},
\eq
where $a_1, \; t_1$ and $\rho_1$ are the integration constants, and the ``-" (``+") sign  corresponds to the contracting $\dot{H} < 0$ (expanding  $\dot{H} > 0$) phase of the universe. Again, without loss of the generality, we can always set $t_1 = 0$, so the
universe is contracting for $t < 0$ and expanding for $t > 0$. At $t = 0$ we have $a = 0$ and the universe evolutes from its contracting phase into an expanding one. However, the singularity at $t =0$ makes the two phases classically 
disconnected.

In order to understand dynamics via the modified FR equations,  
 it is useful to note the 
behavior of the momentum $b$ as shown in  Figs. \ref{fig2} and \ref{fig3}: 
it resembles a step function which changes its value abruptly near the bounce, while far from the bounce
the function $b$ asymptotically approaches two different constant values $b_{\pm\infty}$. In fact, the difference between LQC and the LQG cosmology in the pre-bounce branch lies in the values of the momentum $b$ when $t \rightarrow - \infty$. 
In LQC, $b$ tends to $b_{-\infty}^{{LQC}} \simeq
 1.38176$ while in the LQG cosmology, it goes to $b_{-\infty}^{{LQG}} \simeq 0.588321$. These values are directly determined 
 by the corresponding Hamiltonian constraints in LQC and LQG cosmology. In LQC, the Hamiltonian constraint takes the form 
\bq
\lb{lqcA}
\sin\left(\lambda b\right)=\pm \sqrt{\frac{\rho}{ \tilde \rho_{c}}}.
\eq
 Now,  if one starts from a small energy density in  the post-bounce phase at large volumes and propagates the system backwards in time, the energy density will increase continuously until it reaches 
 its maximum $\tilde \rho_{c}$ at the bounce.  Then,  it will keep decreasing until  
  zero at $t=-\infty$. Choosing the ``+" sign in Eq.(\ref{lqcA}), we find that this corresponds to 
 the case where  $\sin(\lambda b)$ increases from a small value to one at the bounce then drops again to zero, which gives us the limit of the momentum $b$, that is, $b\rightarrow \pi/\lambda= 1.38176$ when $t\rightarrow -\infty$.

\begin{figure}
{
\includegraphics[width=8cm]{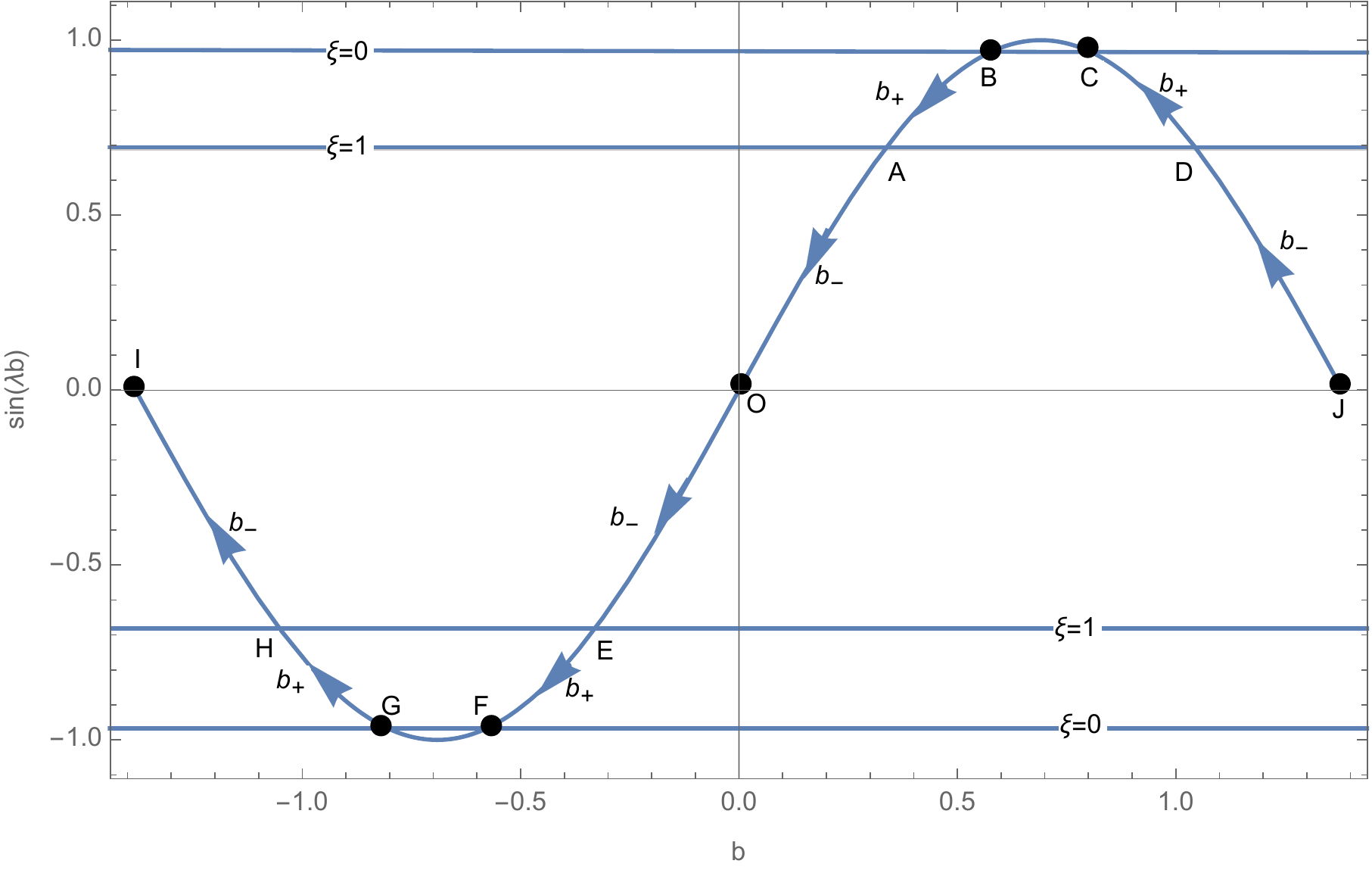}
}
\caption{In this figure, we plot  a complete cycle of $\sin(\lambda b)$ in the interval $\lambda b\in (-\pi, \pi)$ where $\xi\equiv\rho/\rho_c$. The four straight lines $\xi=0$ and $\xi=1$ are where 
$\sin(\lambda b)= \pm \sqrt{1/{(\gamma^2+1)}}$ and $\sin(\lambda b)= \pm \sqrt{1/{[2(\gamma^2+1)]}}$, respectively. The black dots on the curve represent the starting points (G, O, B, J) 
or endpoints (C, O, F, I) of the evolution,  which takes place in the direction indicated by  the arrows. The segments without the arrows ($C\rightarrow B$ and $F\rightarrow G$) are the forbidden 
regions where the energy density exceeds the critical density $\rho_c$. In our numerical simulations, we focus on the line $B\rightarrow A\rightarrow O$.   }
\label{fig6}
\end{figure}

\begin{figure}
{
\includegraphics[width=8cm]{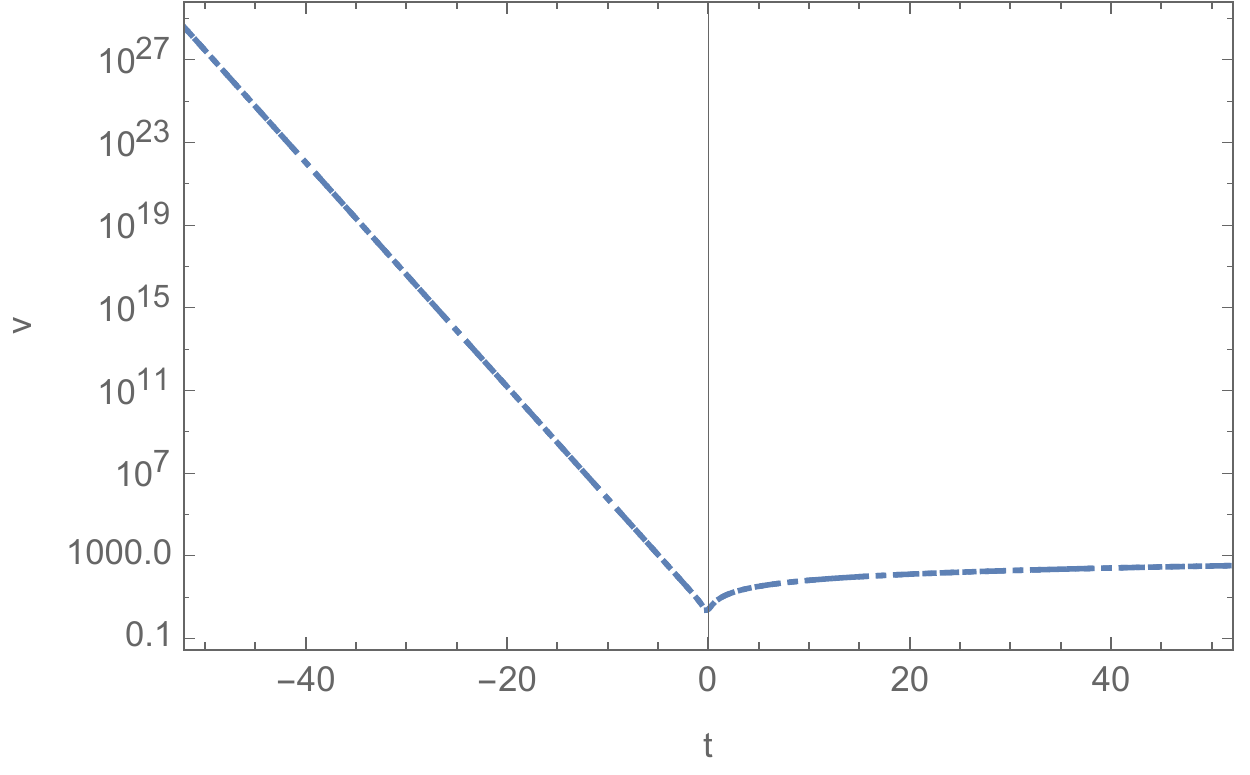}
\includegraphics[width=8cm]{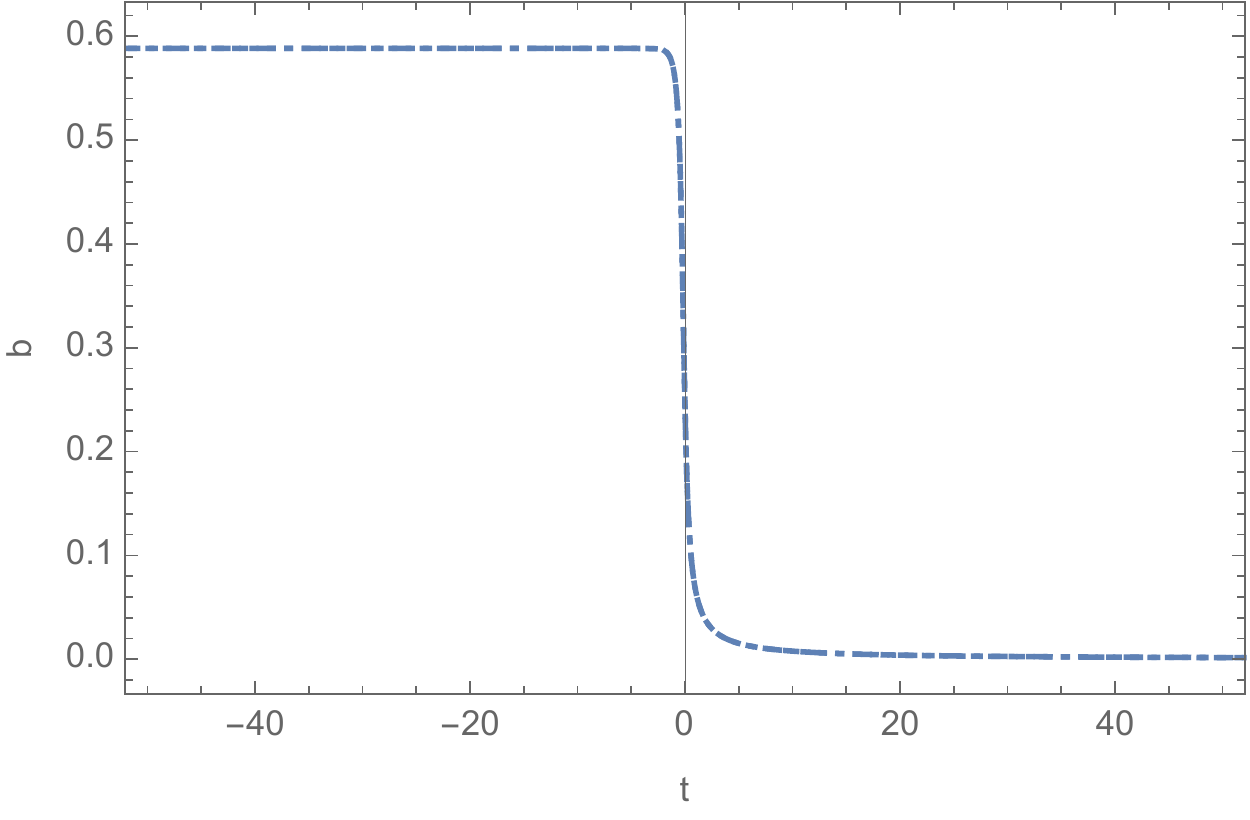}
\includegraphics[width=8cm]{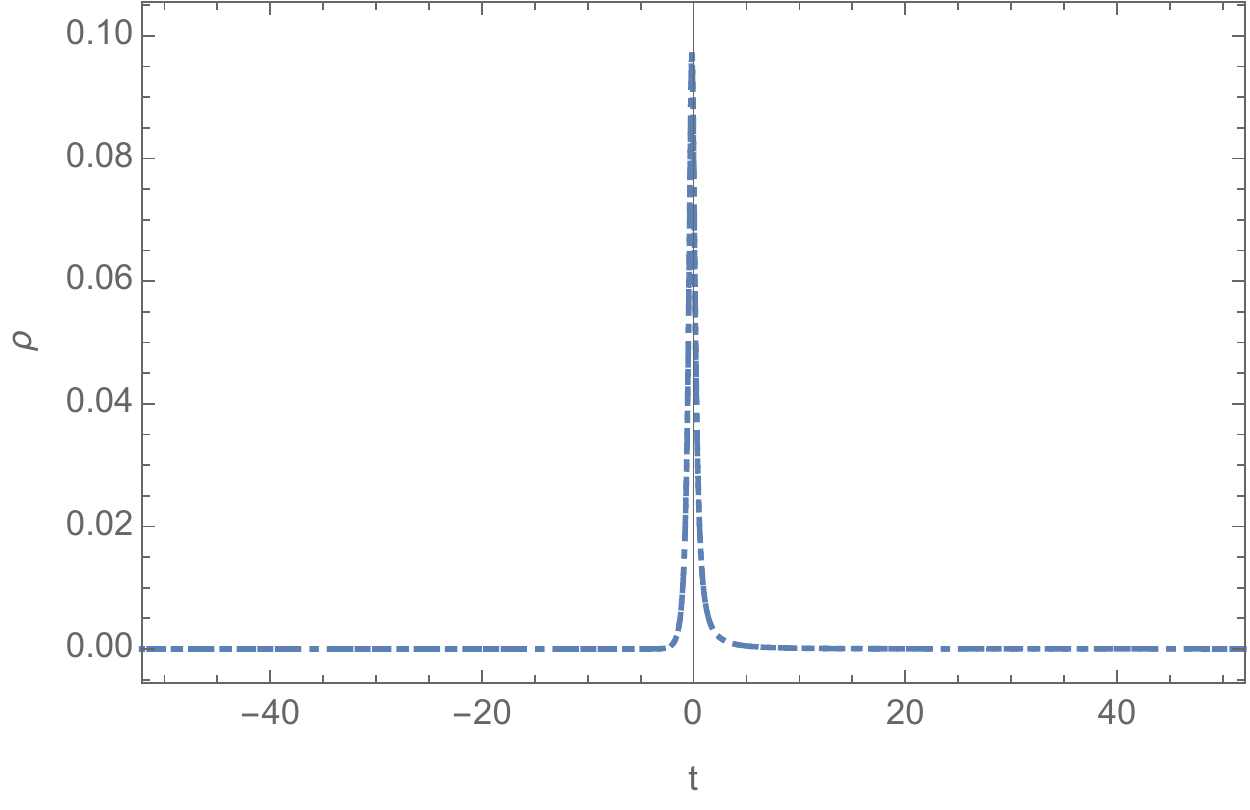}
}
\caption{In this figure, we compare the numerical simulations of the Hamilton's equations (represented by the dotted line) and the combination of the two branches $b_-$ and $b_+$ (represented by the dot-dashed line )
 in the full LQG cosmology, as explained in detail in  Sec. IV.  Our ``initial" data are chosen at  $t_0=115$ with $\rho_0=10^{-6}$, $\pi^0_\phi=1$. The bounce occurs at $t_B=-0.158$.}
\label{fig7}
\end{figure}

Similarly, one can understand the limit of $b$ which tends to $0.588321$ when $t\rightarrow -\infty$ in the LQG cosmology.  Again, the key point is the Hamiltonian  constraint given by 
Eq.(\ref{Hcd}), from which we find,
\bqn
\lb{Hcdb}
\sin(\lambda b_{\pm})= \epsilon_1\sqrt{\frac{1\pm\sqrt{1-\rho/\rho_c}}{2(\gamma^2+1)}},   
\eqn
where $\epsilon_1 = \pm 1$, which also appears in Eq.(\ref{Hcd0}).
In Fig. \ref{fig6}, we plot the function $\sin(\lambda b_{\pm})$ for $\epsilon_1 = \pm1$ and $\lambda b \in (-\pi, \pi)$  to illustrate the evolution of the system governed by the FR equations  (\ref{1.4a})-(\ref{1.4b}) and (\ref{1.6a})-(\ref{1.6b}).  
As indicated by Eqs.(\ref{Hcd1})-(\ref{Hcd2}), for the $b_- $ branch, $|\sin(\lambda b)|\le \sqrt{1/{[2(\gamma^2+1)]}}$,  which
are    the regions labeled with $b_-$ on the curve. Similarly, the lines labeled with $b_+$ are the $b_+$ branch. Since the energy density   approaches zero in the 
distant past and future, there are altogether four distinctive processes implied in Fig. \ref{fig6}:
\bqn
\lb{SGs}
 i) \; J &\rightarrow& D\rightarrow C; \quad ii) \; O\rightarrow E\rightarrow F,\nb\\
 iii) \; B &\rightarrow&  A\rightarrow O,  \quad iv) \; G\rightarrow H\rightarrow I.
\eqn
The first two processes can be immediately ruled out  since,  as already pointed out in Sec. II, only the $b_-$ branch has the classical limit in the post-bounce phase which is consistent with current observations. 
If one starts from a small positive $b$, and evolve backwards in time as done in the above simulations, the universe evolves by following the third segment $BAO$  
of Eq.(\ref{SGs}) backward, as shown in Fig. \ref{fig6}. 
During the part $O\rightarrow A$, the dynamics is described by the $b_-$ branch, i.e. the FR equations (\ref{1.4a})-(\ref{1.4b}).  Then, at the bounce (the point A where $\rho = \rho_c$),  the $b_+$ branch  
takes over [cf. Fig. \ref{fig1}], and the universe follows the evolution given by
Eqs. (\ref{1.6a})-(\ref{1.6b}) during the phase  $A\rightarrow B$. Finally, when $t\rightarrow -\infty$, $\sin(\lambda b_{+})\rightarrow \sqrt{1/{(\gamma^2+1)}}$ and $b_+(t) \rightarrow 0.588321$, as shown in Figs. \ref{fig2} and \ref{fig3}.

It is important to note that in the LQG cosmology, there are two different sets of the modified FR equations corresponding to post-bounce phase (\ref{1.4a})-(\ref{1.4b}), and pre-bounce phase (\ref{1.6a})-(\ref{1.6b}). Starting from the pre-bounce or the post-bounce phase with one of these sets of equations, a consistent evolution is achieved only when the switch over to the other set of equations is performed at the bounce. As a result an asymmetric bounce is inevitable. If one naively chooses only one set of equations,  corresponding to one of the roots from $b_+$ and $b_-$, and assumes its validity at all times both before and after the bounce, one will find that  the trajectory is symmetric. However, it is easily checked that the Hamiltonian constraint is not satisfied in such an evolution as we expect from the above arguments. In other words, a symmetric solution is physically inconsistent in the LQG cosmology for the spatially flat FLRW model.

It should be noted that in LQC the situation is quite different from what we described above in the full LQG cosmology. In particular, in LQC there is only one set of the modified FR equations (\ref{3.8ca}--\ref{3.8cb}), which are valid 
for all times and yield just one branch solution for the evolution of the universe. The resulting evolution  is symmetric with respect to the bounce point. The Hamiltonian constraint is satisfied for this symmetric solution.
  
In Fig. \ref{fig7}, we compare the numerical results of the Hamilton's equations with those obtained from the combination of the FR equations (\ref{1.4a})-(\ref{1.4b}) and  (\ref{1.6a})-(\ref{1.6b}). 
Initially (again recall we are integrating the systems backwards), we start with the $b_-$ branch by using Eqs.(\ref{1.4a})-(\ref{1.4b}) until the  bounce $\rho = \rho_c$ when $|\sin(\lambda b_-)|=\sqrt{1/{[2(\gamma^2+1)]}}$. Then, taking the values at the bounce as the  initial 
conditions for the  $b_+$ branch, we integrate  Eqs.(\ref{1.6a})-(\ref{1.6b}) backwards in time. Note that at the bounce, both of the sets of the modified FR equations agree with each other, the solutions match and are continuous. The resulting evolution yields dynamics which is identical to the one obtained from the Hamilton's equations. In this way, we find the results obtained from the two different approaches agree with each other within the allowed numerical errors. 

It must be noted that, although the conclusion is  well expected, the above identifications are non-trivial. In particular, we find that  the Hamilton's equations (\ref{eomA})-(\ref{eomB}) with the effective Hamiltonian 
constraint (\ref{HCD}) uniquely determine the evolution of the universe, once the initial conditions are given. But, this unique trajectory of motion is represented by two different branches of the FR 
equations, always switched one from the other   at the bounce as shown explicitly by Fig. \ref{fig1}. In particular, the matching
$({\mbox{pre-bounce, post-bounce}}) =  (+, +)$ or $({\mbox{pre-bounce, post-bounce}}) =  (-, -)$ are inconsistent with  the Hamilton's equations, while the matching $({\mbox{pre-bounce, post-bounce}}) 
=  (-, +)$ is ruled out by the current observations. (Here 
``$\pm$" refer to the $b_{\pm}$ branches). Therefore, the only possible matching is the one $({\mbox{pre-bounce, post-bounce}}) =  (+, -)$, which is consistent with both the Hamiltonian evolution 
and the current observations.

\section{Massive Scalar Field }
\label{Section5}
\renewcommand{\theequation}{5.\arabic{equation}}\setcounter{equation}{0}

  \begin{figure}
{
\includegraphics[width=8cm]{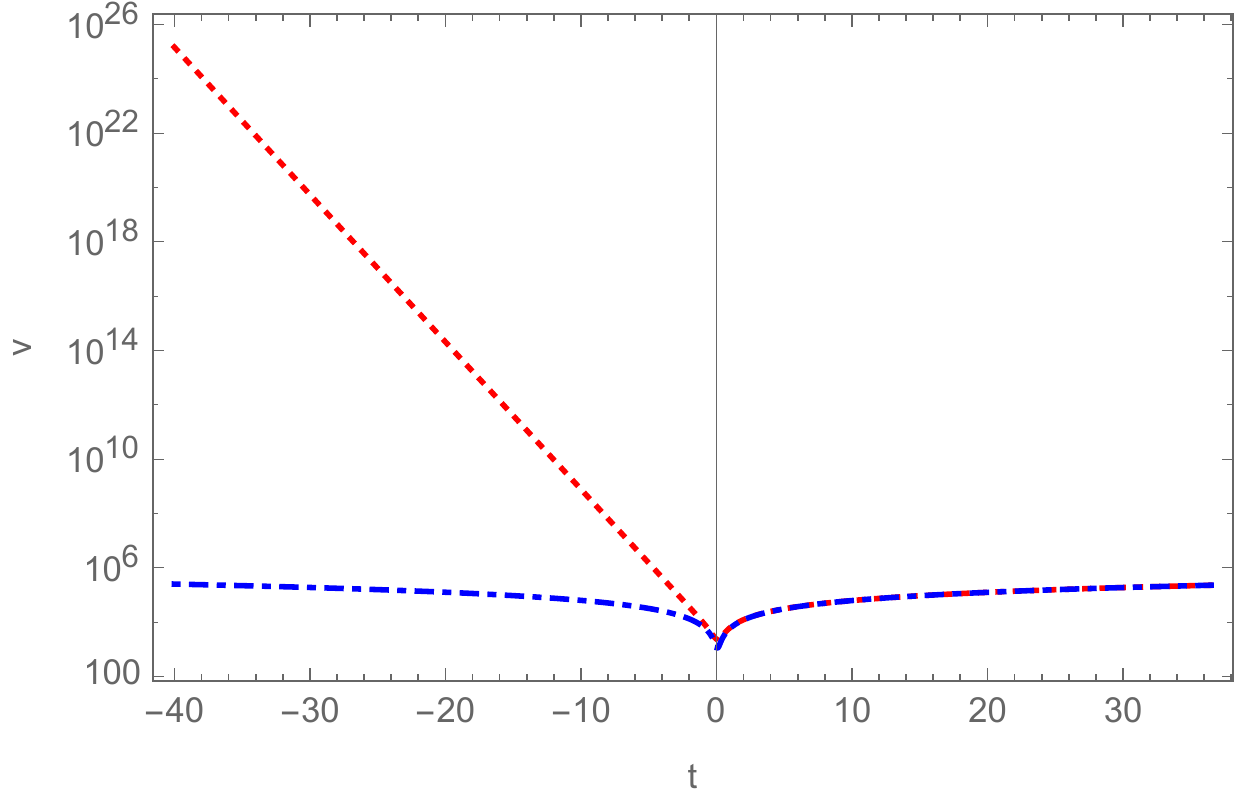}
\includegraphics[width=8cm]{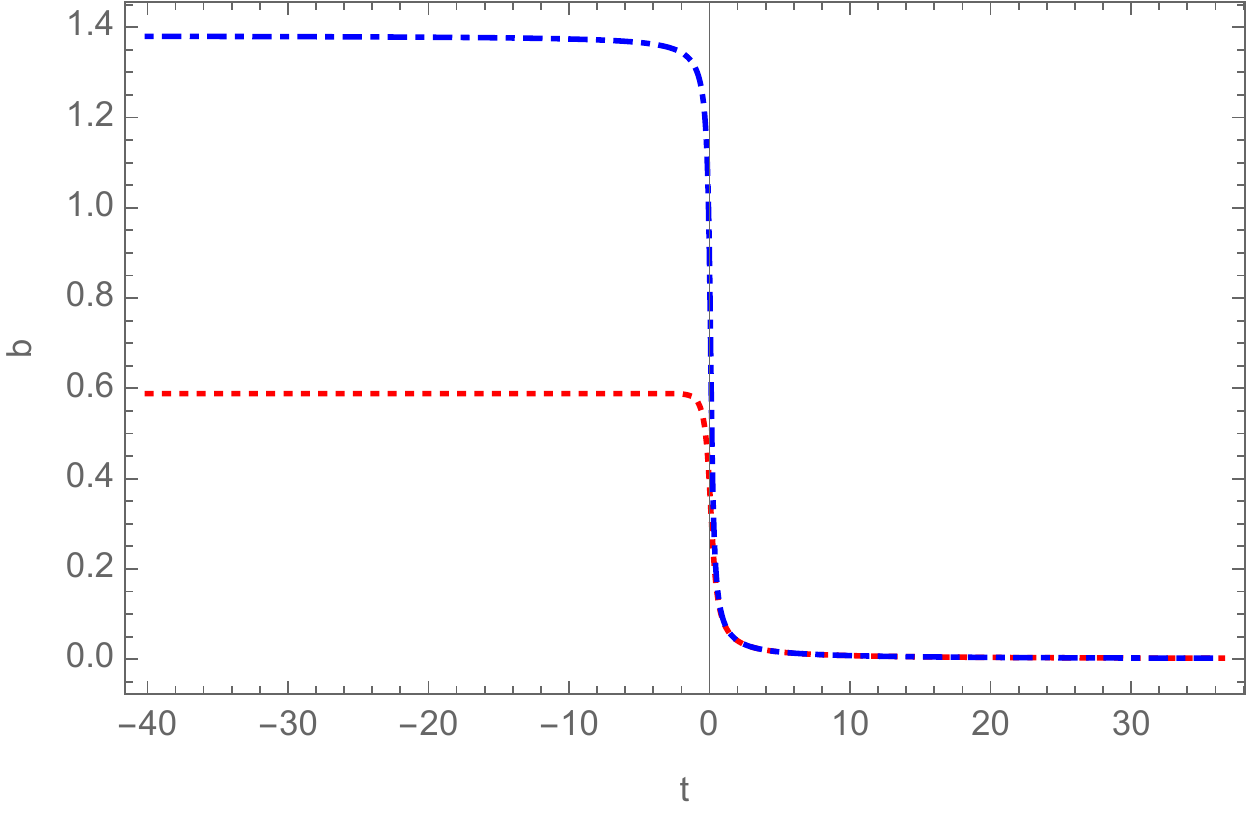}
\includegraphics[width=8cm]{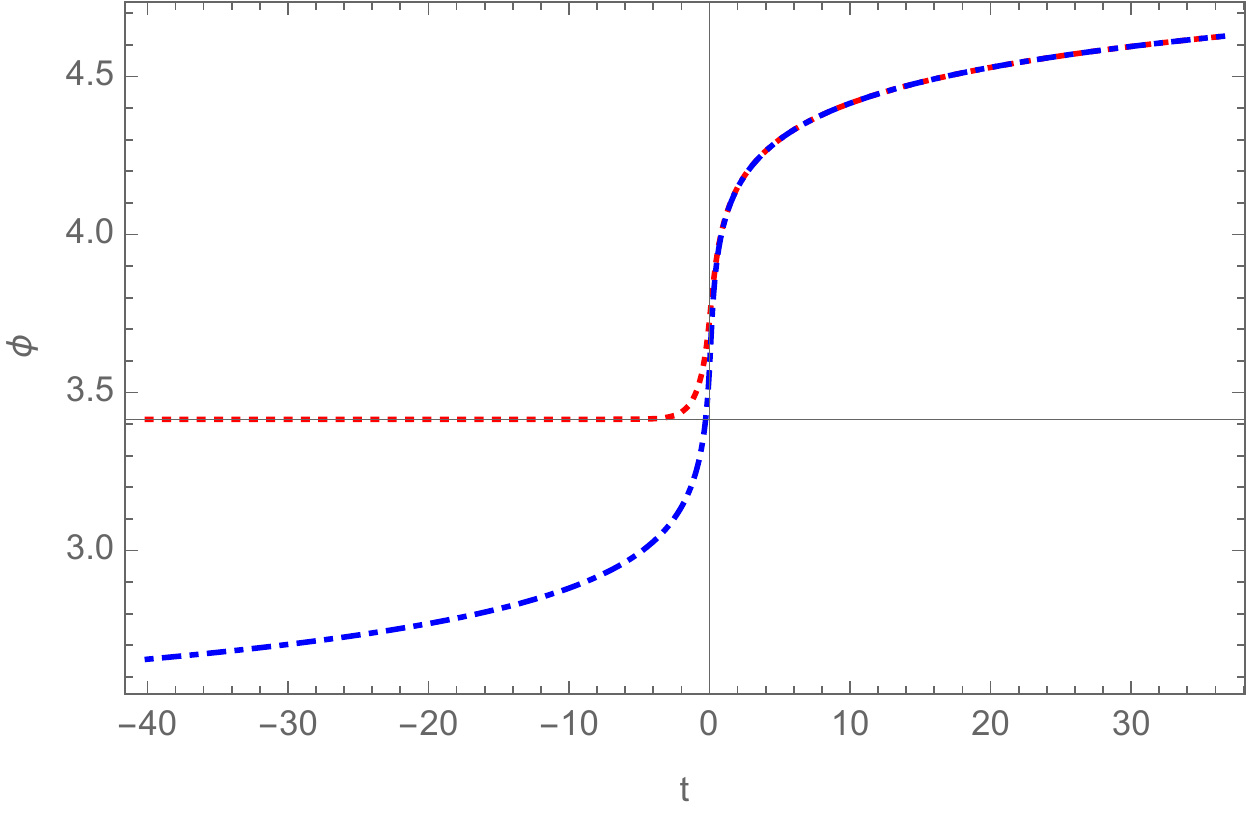}
}
\caption{In this figure, we show the numerical simulations with a massive scalar field and the mass is set to $m = 1.3\times 10^{-6}$.  Our ``initial" conditions are chosen at  
$t_0=36.5$ with $\rho_0=10^{-5}$, $\pi^0_\phi=1000$ and  $v_0=223607$ so that $\phi_0=4.63$. The red dotted lines represent the results from LQG cosmology, and the blue dot-dashed lines are for LQC. }
\label{fig12}
\end{figure}

  \begin{figure}
{
\includegraphics[width=8cm]{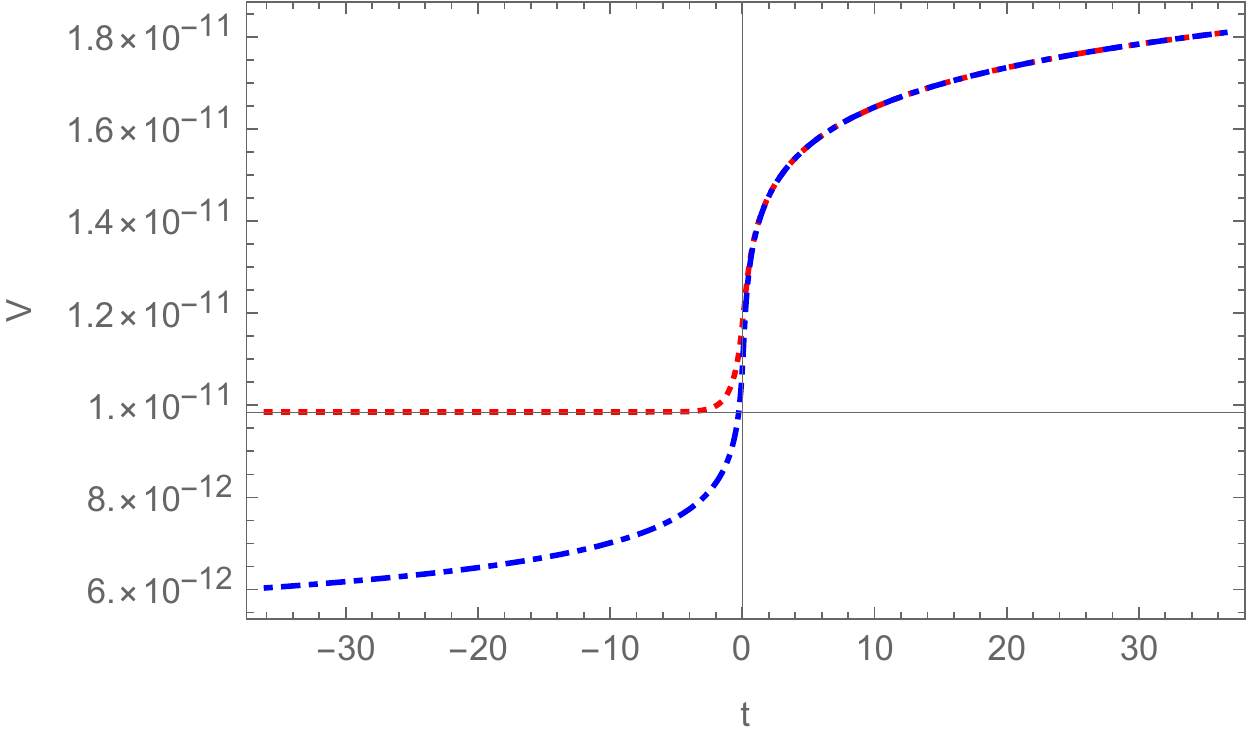}
\includegraphics[width=8cm]{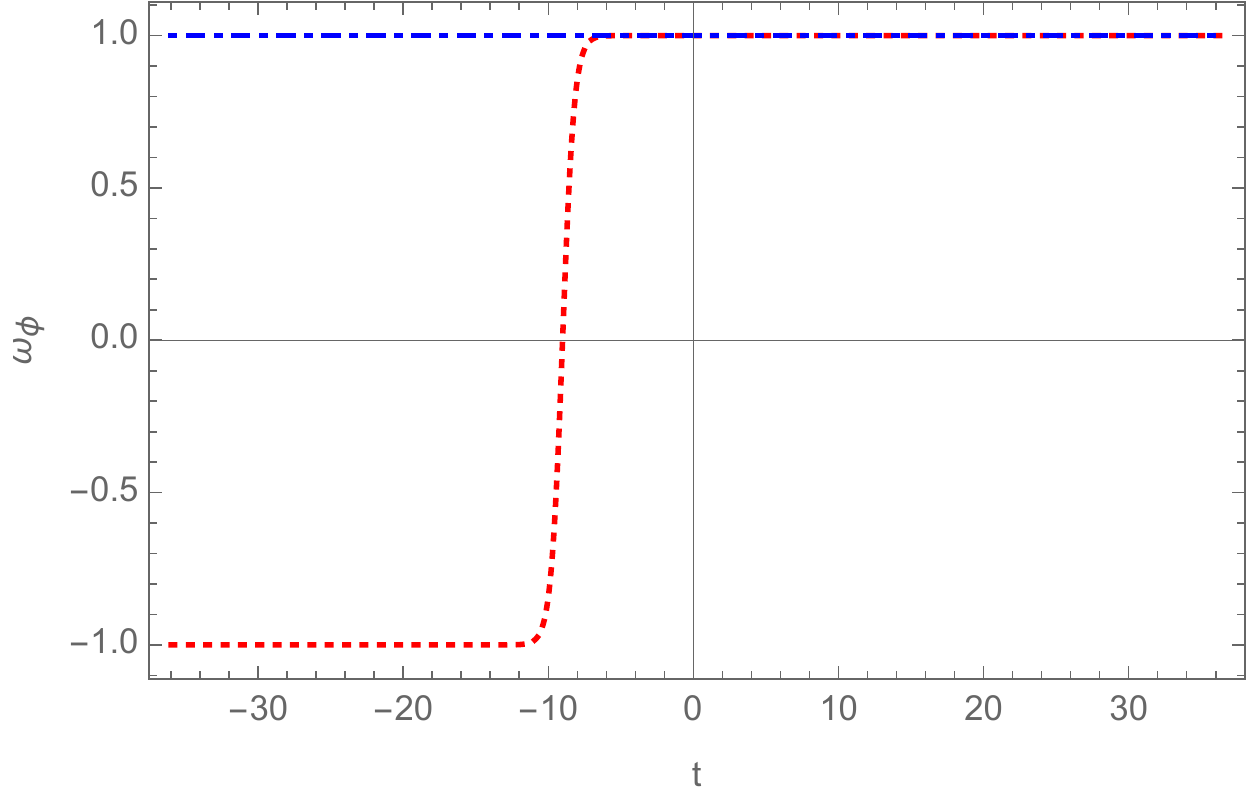}
\includegraphics[width=8cm]{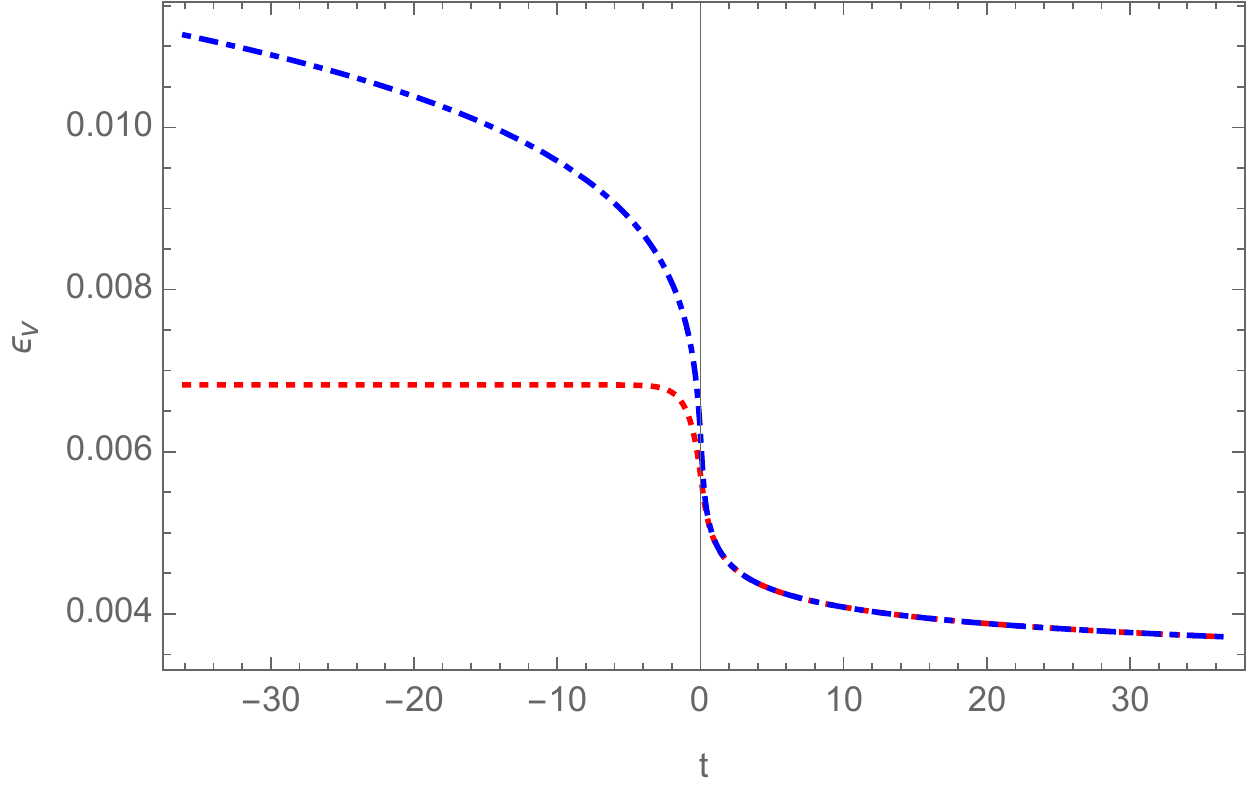}
}
\caption{With the same ``initial" conditions as adopted in Fig. \ref{fig12}, we show the behavior of the potential, equation of state and slow roll parameter near the bounce. The LQG results are shown by 
the red dotted curve, and the LQC results  are depicted by the blue dot-dashed curve.}
\label{fig13}
\end{figure}

   \begin{figure}
{
\includegraphics[width=8cm]{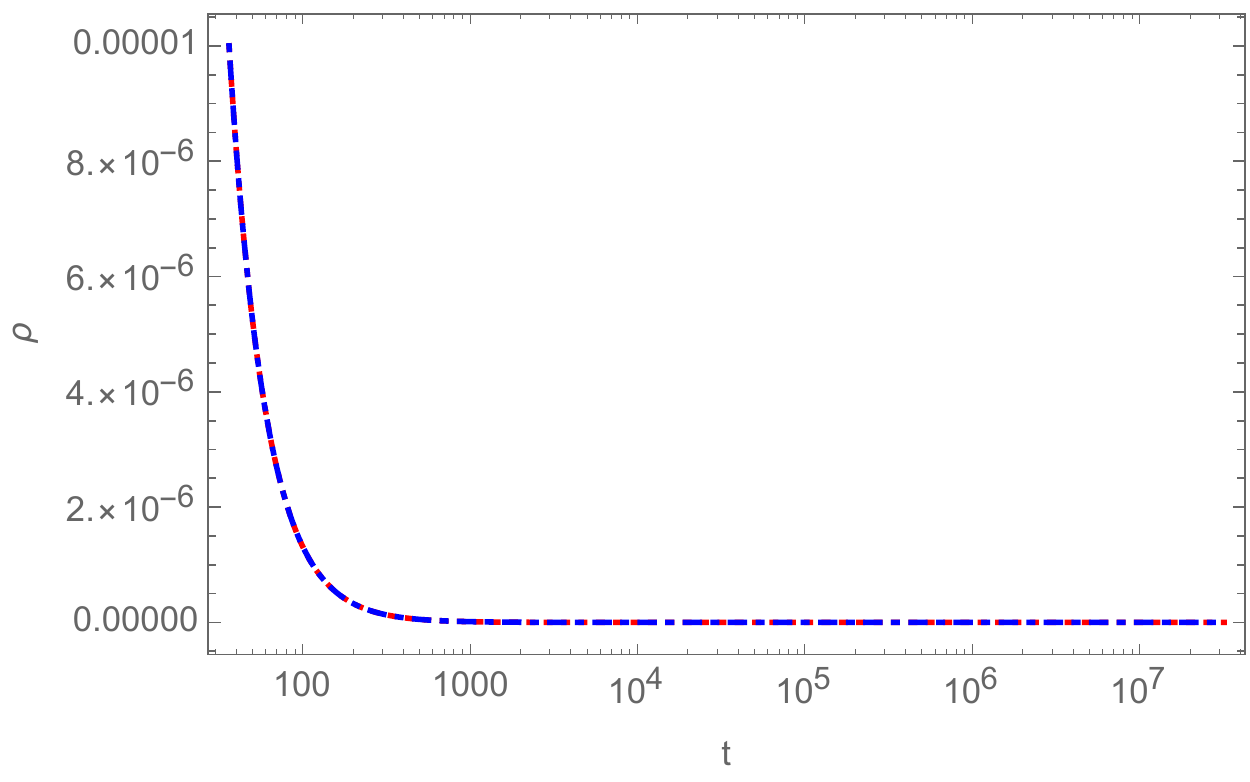}
\includegraphics[width=8cm]{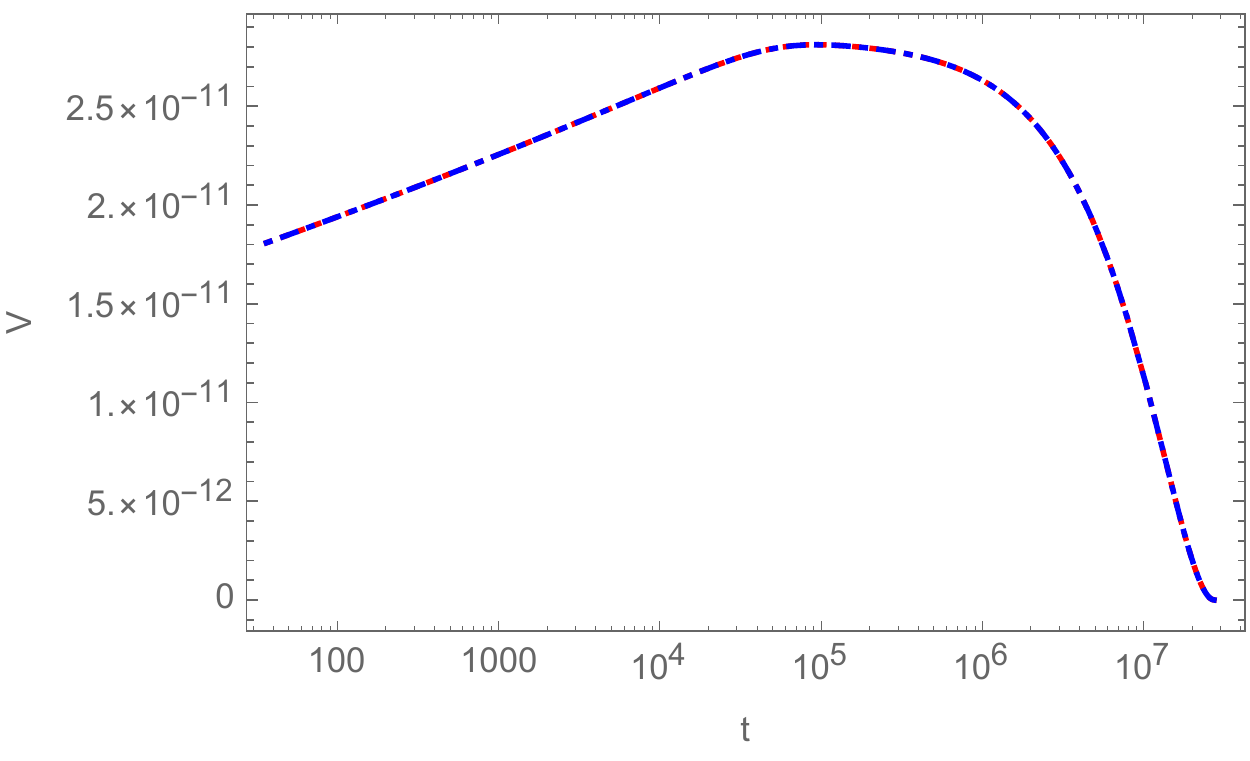}
\includegraphics[width=8cm]{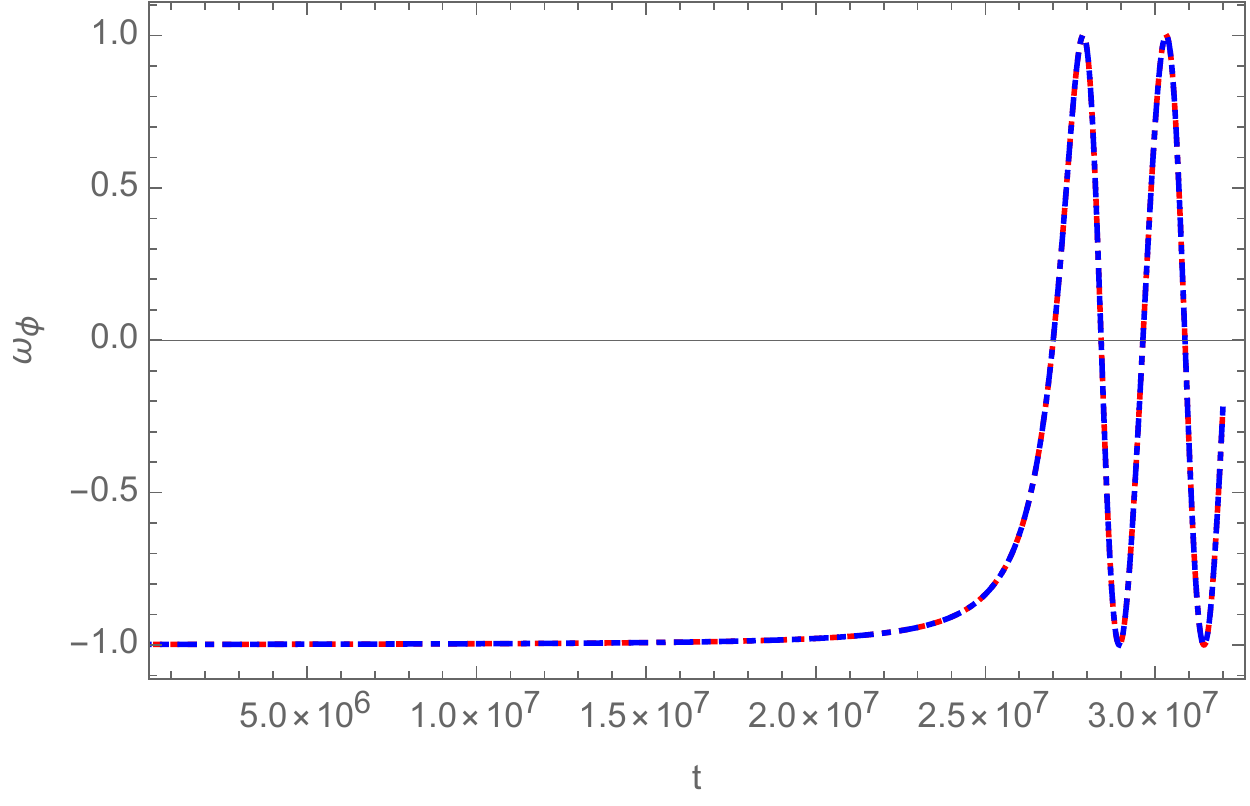}
\includegraphics[width=8cm]{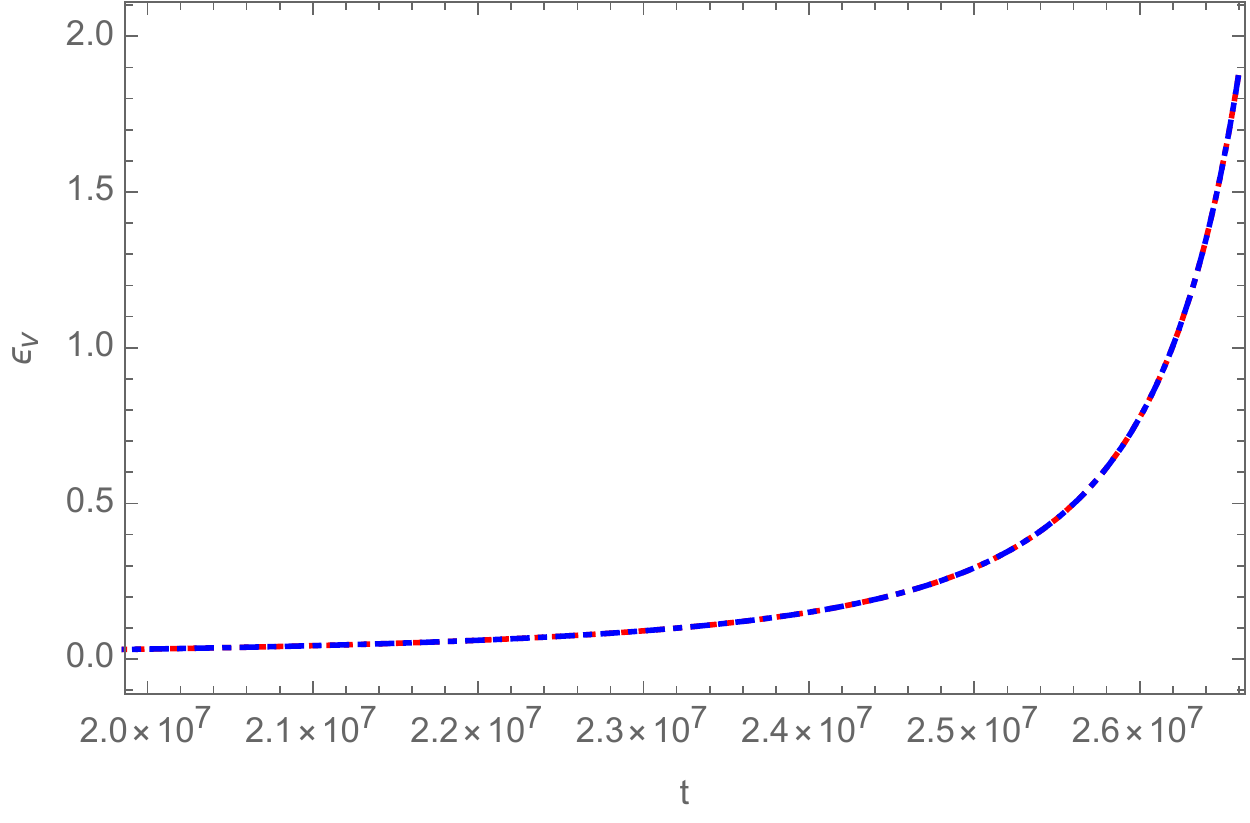}
}
\caption{The plots with  the same initial conditions as those in Fig. \ref{fig13} are shown for late times in the post-bounce branch during the inflationary phase and its end. 
 The slow-roll parameter increases to unity at  $2.6\times 10^7\; t_{Pl}$. }
\label{fig14}
\end{figure}

  \begin{figure}
{
\includegraphics[width=8cm]{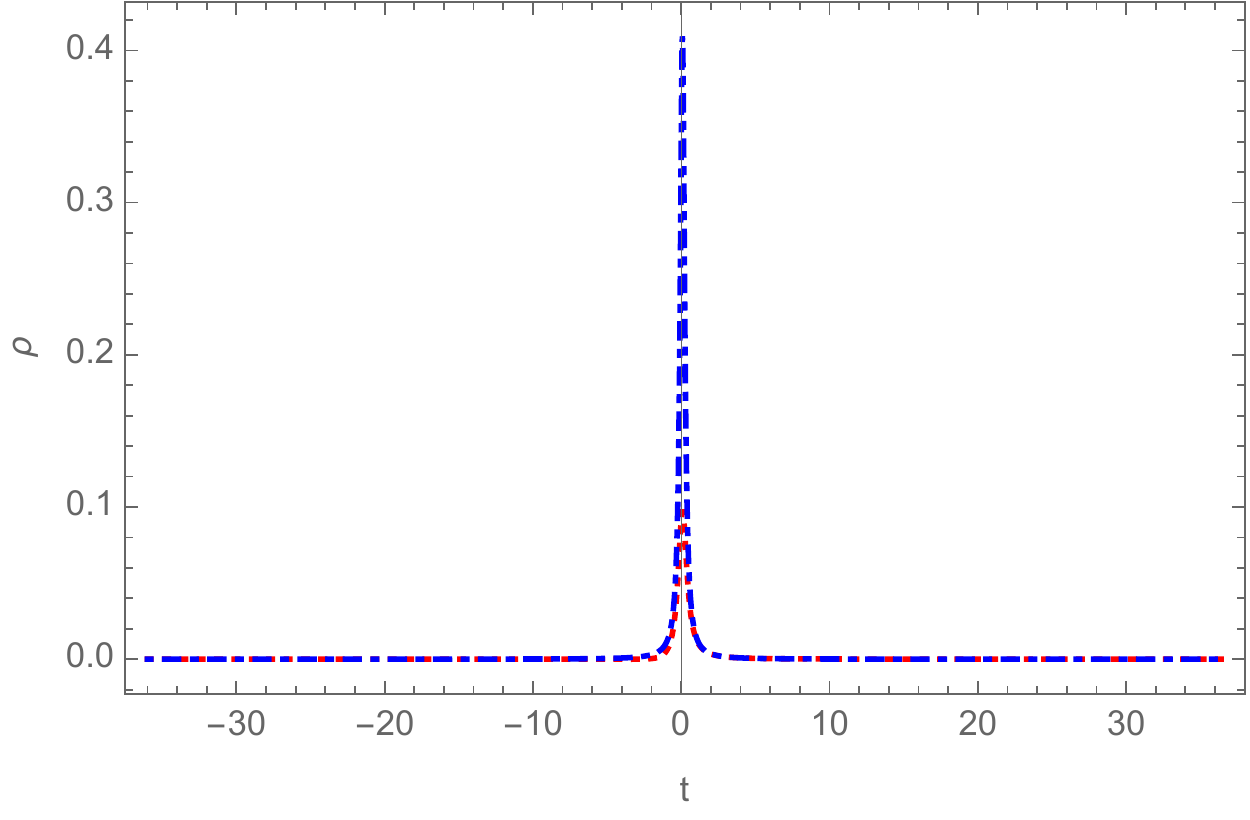}
\includegraphics[width=8cm]{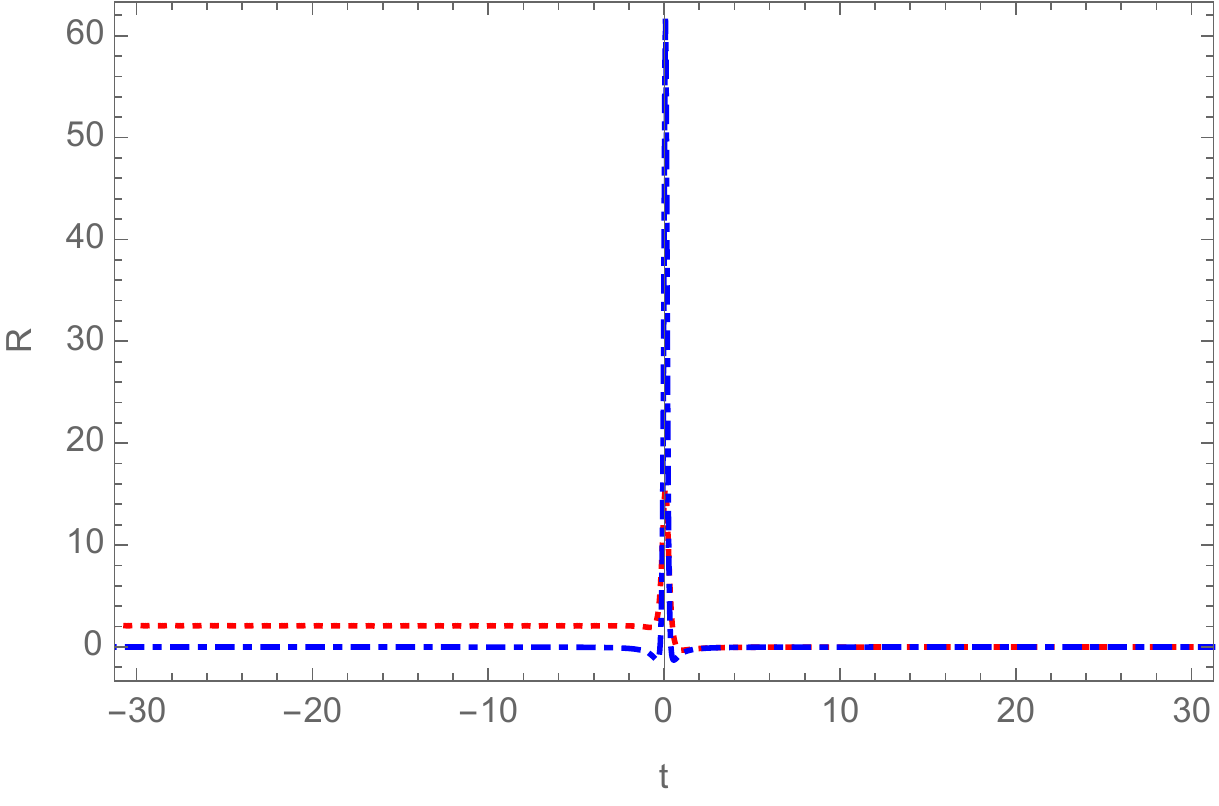}
\includegraphics[width=8cm]{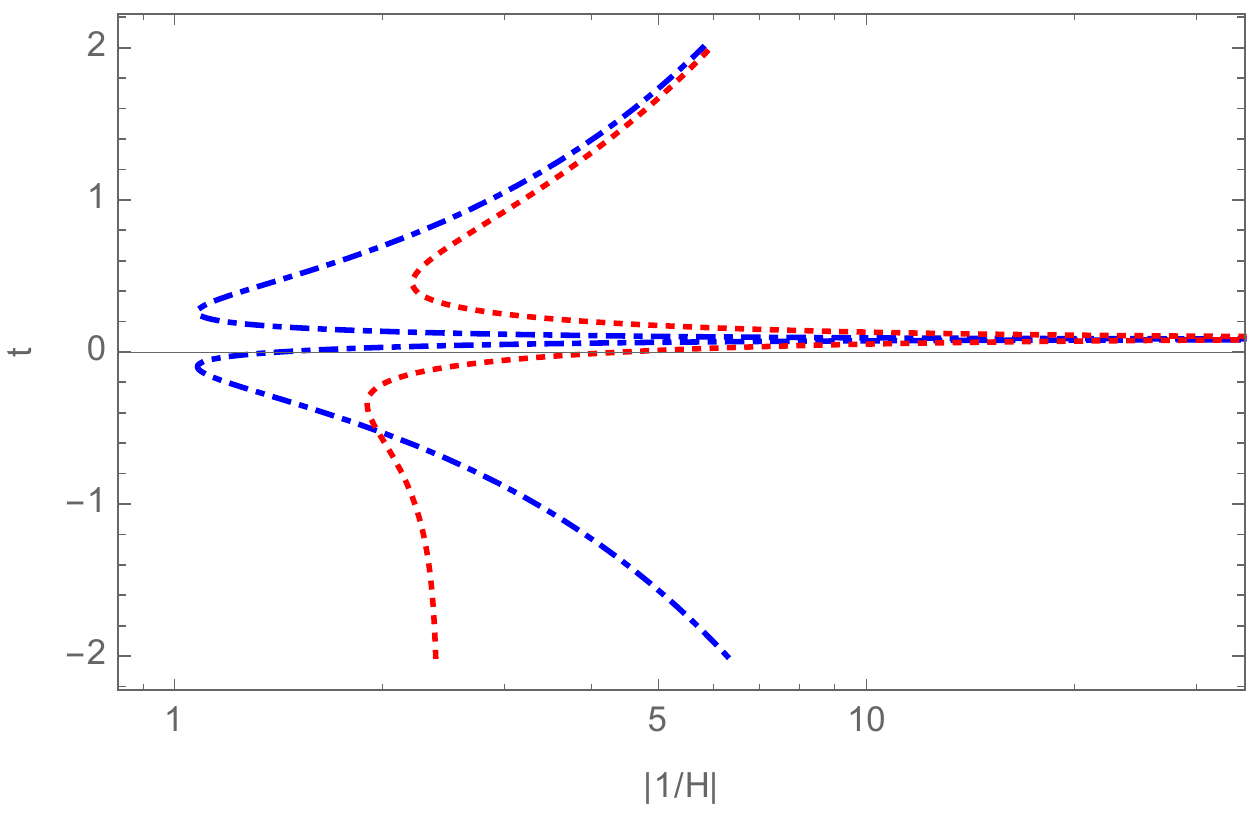}
}
\caption{With the same ``initial" conditions as adopted in Fig. \ref{fig12}, we show the behavior of the energy density, Ricci scalar and inverse Hubble rate near the bounce. The LQG results are shown by 
the red dotted curve, and the LQC results  are depicted by the blue dot-dashed curve. In the third subfigure, we also plot the Ricci scalar in a longer range after the bounce to show that there exists a negative curvature regime in the classical limit.  }
\label{fig15}
\end{figure}

In this section, we will carry out the numerical simulations for the massive scalar field with a quadratic  potential. In this case, the matter Hamiltonian reads:
\bq
\lb{5.1}
\mathcal{H}_M  =\frac{\pi^2_\phi}{2 v}+\frac{v}{2}m^2\phi^2.
\eq
As a result, the Hamilton's equations of the massive scalar field now become 
\bqn
\lb{5.2}
\dot \phi&=&\frac{\pi_\phi}{v},\nb\\
\dot \pi_\phi&=&-vm^2\phi,
\eqn
which lead to the Klein-Gordon equation of the scalar field
\bq
\lb{5.3}
\ddot \phi+3H\dot \phi+m^2\phi=0.
\eq
The energy density and pressure of the massive scalar field are  now given by  
\bqn
\lb{5.4}
\rho&=&\frac{H_M}{v}=\frac{\pi^2_\phi}{2v^2}+\frac{1}{2}m^2\phi^2,\nb\\
P&=&-\frac{\partial H_M}{\partial v}=\frac{\pi^2_\phi}{2v^2}-\frac{1}{2}m^2\phi^2.
\eqn
Its dynamical evolution can be obtained by substituting the above  expressions of $\rho$ and $P$ into the Hamilton's Eqs.(\ref{eomA})-(\ref{eomB}), 
or the modified FR equations (\ref{1.4a})-(\ref{1.4b}) and (\ref{1.6a})-(\ref{1.6b}). Numerical simulations yield the same result once we take into account the switch between two sets of the 
modified FR equations at the bounce.

To understand the resulting dynamics, let us note that  
during inflation (where $\rho \ll \rho_c$), the potential energy of the inflaton starts  dominating, and the Hubble rate is almost constant and can be approximated by
\bq
\lb{5.6}
H^2\approx \frac{8\pi G V(\phi)}{3},
\eq
during which the two slow-roll parameters defined by \cite{DB},
\bqn
\lb{5.8}
\epsilon_V\equiv \frac{M_{pl}^2}{2}\left(\frac{V'}{V}\right)^2,\quad
\eta_V \equiv M_{pl}^2 \left(\frac{V''}{V}\right),
\eqn
 are expected to be very small, $\epsilon_V, \; \left|\eta_V\right| \ll 1$, where $M_{Pl}^2 \equiv 1/(8\pi G)$. Therefore, during the inflationary phase, we  have $w_{\phi}  \simeq -1$, where
 \bqn
\lb{5.8aa}
w_{\phi} \equiv \frac{P}{\rho} = \frac{\dot\phi^2 - 2V(\phi)} {\dot\phi^2 + 2V(\phi)}.
\eqn

In the case of the quadratic potential $V=m^2\phi^2/2$, these two parameters have the same value, and are given by, 
\bq
\epsilon_V=\eta_V=\frac{2M_{pl}^2}{\phi^2}.
\eq
Thus, to have $\epsilon_V, \; \left|\eta_V\right| \ll 1$, we must assume that $\phi \gg M_{pl}$ during the period of inflation. In LQC, to be consistent with  
current observations, and meanwhile allowing non-negligible  quantum gravitational effects, it was numerically estimated that $m\phi_B \simeq 10^{-6}$ \cite{AM15}, where $\phi_B$ is the value of the scalar field at the bounce. While the best
 fitting data of the Planck 2015
 yields  $m \simeq 1.3 \times 10^{-6}$  \cite{Planck2015}.  It should be noted that in obtaining this value of mass $m$, the pre-inflationary 
dynamics was not taken into account \cite{Planck2015}. However, it was shown recently that the quantum 
gravitational effects relax the observational constraints in LQC \cite{Zhu1}. In particular,  the tension between this quadratic potential model and observations 
found in \cite{Planck2015} can be alleviated  by properly choosing the free parameters that characterize  the  quantum  gravitational effects  \cite{Zhu1}. 

 In our analysis, we  simply choose  $m \simeq 1.3 \times 10^{-6}$, as we find that the main properties remain the same for different choices of $m$ for  $m \in \left(10^{-4}, 10^{-7}\right)$.
 In particular,  in Fig. \ref{fig12} we plot the functions $v, \; b,\; \phi$ vs the time $t$. During the whole process of the simulation, we also monitor the numerical errors in validity of the 
 Hamiltonian constraint, which  turn out to be always very small. As a result, our numerical
simulations are quite  reliable and confined well to the constraint surface ${\cal{C}} \approx 0$ within the numerical accuracy. Similar to the massless case, the evolution of the universe before the 
quantum bounce in the full LQG cosmology is quite different from that of LQC in terms of  the three quantities, $v, \; b, \; \phi$, while after the bounce the two theories give rise to similar behavior.
In Fig. \ref{fig13}, we plot the quantities $\rho, \; V(\phi), \; w_{\phi}$ and $\epsilon_V$, from which we can see that the universe is de Sitter-like in the pre bounce phase in the LQG cosmology,
so we have $ w_{\phi} \simeq -1$, as shown by the red dotted lines, while it is completely dominated by the kinetic energy of the inflaton in LQC, so that in LQC we have $ w_{\phi}^{LQC} \simeq +1\; (t \le t_B)$.
Near the bounce, the kinetic energy of the inflaton rises dramatically, so the evolution of the universe in both theories is dominated by it, and, as a result, we have $ w_{\phi} \simeq +1$ near the 
bounce in both theories. Fig. \ref{fig14} shows the long time evolution of the same quantities as shown in Fig. \ref{fig13}. From this figure we can see that a slow-roll inflation results after the quantum
bounce, and the inflation ends at about $t \simeq 2.6 \times 10^{7}\; t_{Pl}$, at which we have $\epsilon_v \simeq 1$ and $ w_{\phi} \simeq -1/3$.

  In Fig. \ref{fig15}, we  also show the behavior of the energy density of matter,  Ricci scalar and inverse Hubble rate near the bounce for both LQC and the LQG cosmology, with  the same ``initial" conditions as adopted in Fig. \ref{fig12}.  As earlier, we find that in the pre-bounce phase the spacetime retains its quantum gravitational character even at early times.

\section{Summary}
\renewcommand{\theequation}{6.\arabic{equation}}\setcounter{equation}{0}

In this paper, we have systematically studied the evolution of the spatially flat FLRW universe using an effective Hamiltonian recently obtained using complexifier
coherent states in LQG \cite{DL17}. The same Hamiltonian was obtained earlier in LQC by keeping the Lorentzian term
separate from the Euclidean one \cite{YDM09}. Recently, in  \cite{adlp} loop quantization of this Hamiltonian was studied in the $\bar \mu$ scheme 
(see \cite{aps3}), and  properties of the physical Hilbert space were investigated including the behavior of eigenfunctions. It was found that 
loop quantum geometric effects result in an effective cosmological constant in one of the branches in evolution. In the latter study, physical states and Dirac observables were constructed 
and numerical simulations were performed using massless scalar field as a relational clock. Numerical simulations showed that big bang 
singularity is resolved and states remain sharply peaked throughout the evolution. Further,
effective dynamics as captured from the Hamilton's equations was shown to be a good approximation to the underlying quantum dynamics. 
However, so far only preliminary results of the effective dynamics were available in all these studies. In particular, the modified FR equations were not known. 
Further, existing numerical studies in both set of works, \cite{YDM09} and  \cite{DL17,adlp}, were not in harmony with each other, with the latter claiming an asymmetric bounce with a Planckian 
curvature de Sitter phase in one side of the bounce, and the former indicating a symmetric bounce with a pre-bounce and post-bounce phases
asymptotically approaching a small curvature spacetime described by GR.
In addition,  \cite{DL17,adlp} did not addressed some important questions, including the exact cause of the disagreement with earlier results from \cite{YDM09}, what 
is the modified Friedmann dynamics and how it differs from LQC, and the reason of the asymmetric bounce versus the symmetric bounce in \cite{YDM09}.  The  results from the latter, if  true,  bring  the
effective dynamics in the LQG cosmology very close to the  picture  in  LQC  where  for  the  spatially  flat  FLRW
model  the  evolution  of  the  universe  is  symmetric  with respect  to  the  moment  of  bounce  \cite{aps3}.   However,  if  the
bounce is asymmetric, then there are qualitative differences between LQC and the considered LQG cosmology,
especially  in  the  pre-bounce  phase  which  can  result  in potentially significant phenomenological differences. Thus, studies so far, including  \cite{YDM09,DL17,adlp}, 
resulted in a tension on resulting physics from the LQC Hamiltonian where Lorentzian term is treated independently. The 
reason for this tension was that various subtleties of the effective dynamics were not well known and understood in this model.

 While the analysis in \cite{adlp} addressed the quantum treatment of the Hamiltonian, our analysis focused on the details 
and subtleties of the effective dynamics. In particular, our study filled important gaps on the understanding of the origin of 
asymmetric bounce. Unlike previous studies we found that not only loop quantum effects result is an emergent cosmological 
constant but also to a modified Newton's coupling constant.  The main goal of this paper was to understand the effective dynamics 
in detail and obtain a consistent picture of the singularity resolution in this LQG cosmology.  In particular, we investigated singularity resolution for both the massless scalar field and the scalar
field with a quadratic potential,  and performed various numerical simulations using the Hamilton�s equations as well as the modified FR equations.

Assuming the validity of the effective Hamiltonian obtained in \cite{YDM09,DL17} throughout the evolution, we first derived the Hamilton's equations and then the modified FR equations in the LQG cosmology. Some surprises appear in comparison to LQC at this level. In LQC, as in GR, there is only one set of the modified FR equations valid for the entire evolution of the universe. The mapping between the Hamilton's equations and the modified FR equations is simple and one-to-one. However, we found that there are two sets of the modified FR equations that are equivalent to the Hamilton's equations in the LQG cosmology. None of these two sets can cover
 the complete evolution of the universe. In particular, the equivalence of the evolution of the universe between the Hamilton's equations and the two sets of equations requires the switch-off of the two branches at the quantum 
 bounce $\rho = \rho_c$. Unlike LQC and GR, there is no one-to-one map between the Hamilton's equations and the modified FR equations in the LQG cosmology. This subtlety is the cause of the existing confusions about the nature of the bounce in this model. Notably the two sets of equations have different asymptotic limits. The set of Eqs.(\ref{1.4a}) -(\ref{1.4b}), which corresponds to the $b_-$ branch in Fig. \ref{fig1}, results in a large universe at late times which has a small spacetime curvature. This is only set with the small spacetime curvature general relativistic limit. On the other hand, the set of the equations (\ref{1.6a}) -(\ref{1.6b}), corresponding to the $b_+$ branch in Fig. \ref{fig1}, results in a de Sitter phase with a Planckian curvature. Interestingly, such an evolution has been found earlier in LQC but for the case of a quantization of  the Schwarzschild interior \cite{djs}. Another contrast with LQC is that 
 the modified FR equations have corrections of higher order, in contrast to  the usual $\rho^2$ corrections in LQC. In another interesting similarity, such corrections were studied earlier in loop cosmology and a similar multiple branch behavior was found \cite{singh-soni}. Notably, one important difference with LQC is the asymmetric super-inflationary regime which is a non-trivial function of the Barbero-Immirzi parameter. The super-inflationary phase in the pre-bounce era is longer than the post-bounce phase. The similarities and contrasts of LQG cosmology with LQC are indeed quite rich and intriguing. 
 
For the dynamical evolution in the LQG cosmology, we found that  the evolution of the universe after the bounce can be asymptotically either de Sitter spacetime or that of GR \footnote{Note that in \cite{DL17,adlp} only the case where
the universe in the post bounce is asymptotically   GR was presented.}. But, the  evolution of the universe with respect to the bounce point is always
asymmetric, so it can only be one of the following two possibilities,
\bqn
\lb{5.1}
 && ({\mbox{pre-bounce, post-bounce}}) 
=  (-, +), \nb\\
 && ({\mbox{pre-bounce, post-bounce}}) 
=  (+, -), 
\eqn
where ``$+$" refers to the asymptotically de Sitter spacetime and ``$-$" refers to the asymptotically GR spacetimes, which correspond to, respectively, the $b_{\pm}$ branches of solutions, as shown in Fig. \ref{fig1}. In this paper, we showed that consistency of the 
Hamiltonian evolution requires that  the combinations ($+, +$) and ($-, -$) are forbidden (see Fig. \ref{fig6}).  If one naively considers only one set of the modified FR equations and evolves through the bounce, one would get an unphysical ``symmetric bounce." Moreover,  the current observational constraints already rule out the 
possibility of the combination ($-, +$).  
Thus,  there is only one possibility for the dynamical evolution in this LQG cosmology. The evolution of the universe must be described by  the set of Eqs.(\ref{1.6a}) -(\ref{1.6b}) before the bounce, and at the bounce it will be switched to  the set of Eqs.(\ref{1.4a}) -(\ref{1.4b}), in order to be consistent with observations.

 All the above were first done for a massless scalar field for which we have $\rho = P$, and then generalized to the case of a quadratic potential. After checking with various simulations, we reached the same conclusions in both  cases on the consistency and equivalence of the Hamilton's equations and the modified FR equations. In particular, we showed that the inflationary phase in the post-bounce regime is compatible with the LQG cosmology where the pre-bounce phase is dictated by the Planck scale cosmological constant. The slow-roll inflation occurs at $t/t_{Pl} \simeq \times 10^{5}$  after bounce and lasts  until
 $t/t_{Pl} \simeq 2.6\times 10^{7}$, which is the same as  found in LQC for the same inflationary  potential \cite{Zhu2,Shahalam}. 
  
 We again emphasize that 
what we have studied in this paper is just one proposal of LQG cosmology and one should be open to  other possibilities in LQG cosmology due to ambiguities in the Hamiltonian constraint. Further, these results were obtained by assuming the validity of the effective dynamics which had an extraordinary success in LQC \cite{aps3,numerics}, but rigorous studies need to be taken in LQG cosmology. With these caveats, we have shown that in the effective Hamiltonian for the spatially flat FLRW spacetime obtained via one of the proposals to regularize the Hamiltonian constraint in LQG,  the big bang singularity is replaced by a big bounce as in LQC. This has been demonstrated for the massless as well as massive scalar field. But there are important distinctions from spatially flat isotropic LQC because the bounce is asymmetric with a Planck curvature de Sitter phase in the pre-bounce branch. Therefore, the problem of the  big bang singularity in GR is resolved even in the full LQG cosmology, and for a massive scalar field,  the slow-roll inflation is always a final result of the evolution of the universe by properly choosing
the initial conditions in the deep Planck era. The modified FR equations which we found in this paper in the LQG cosmology have a very rich structure and serve as a platform for many interesting studies of the Planck scale physics in LQG. For example,  
 it is natural to ask: how natural is it for the slow-roll inflation to happen in the framework of the full LQG cosmology? Are the corresponding linear perturbations consistent with observations?
What are the non-Gaussianity  \cite{ Agullo,Zhu3}? And more importantly, what are the observational signatures and how do they differ from predictions of LQC or other theories of quantum gravity?   We hope to return to these important issues soon.

\section*{Acknowledgements}

We would like to thank Y. Ma, T. Pawlowski, S. Saini, J. Yang and T. Zhu for valuable comments.  A.W. and B.F.L. are supported in part by the National Natural Science Foundation of China (NNSFC) 
with the Grants Nos. 11375153 and 11675145. P.S. is supported by NSF grants PHY-1404240 and PHY-1454832.


\begin{thebibliography}{99}

\bibitem{reviewlqg} C. Rovelli,
{\it{Quantum Gravity}}, Cambridge University Press,  (2004); A.~Ashtekar and J.~Lewandowski,
  Class.\ Quant.\ Grav.\  {\bf 21}, R53 (2004); T. Thiemann,
{\it{Modern  Canonical  Quantum  General  Relativity}}, Cambridge University Press (2007),
K.~Giesel,
  The Kinematical Setup of Quantum Geometry: A Brief Review, in 
{\it{Loop
Quantum Gravity: The First 30 Years}}, Eds: A. Ashtekar, J. Pullin, World Scientific (2017)
  arXiv:1707.03059; A. Laddha and M. Varadarajan, Quantum Dynamics, in 
{\it{Loop
Quantum Gravity: The First 30 Years}}, Eds: A. Ashtekar, J. Pullin, World Scientific (2017).

\bibitem{aps} A.~Ashtekar, T.~Pawlowski and P.~Singh,
  Phys.\ Rev.\ Lett.\  {\bf 96}, 141301 (2006). 

\bibitem{aps3} A. Ashtekar, T. Pawlowski and P. Singh, Phys. Rev. {\bf D}74 (2006) 084003.

\bibitem{slqc}  A.~Ashtekar, A.~Corichi and P.~Singh,
  Phys.\ Rev.\ D{\bf 77}, 024046 (2008).

\bibitem{review1}  M. Bojowald, Living Rev. Rel. 11, 4 (2008).

\bibitem{review2} A.~Ashtekar and P.~Singh,
  Class.\ Quant.\ Grav.\  {\bf 28}, 213001 (2011).

\bibitem{vt} V.~Taveras,
  Phys.\ Rev.\ D{\bf 78}, 064072 (2008).

\bibitem{generic} P.~Singh,
  Class.\ Quant.\ Grav.\  {\bf 26}, 125005 (2009);   P.~Singh and F.~Vidotto,
  Phys.\ Rev.\ D{\bf 83}, 064027 (2011); P.~Singh,
  Phys.\ Rev.\ D{\bf 85}, 104011 (2012); S.~Saini and P.~Singh,
  Class.\ Quant.\ Grav.\  {\bf 33}, 245019 (2016); S.~Saini and P.~Singh,  Class.\ Quant.\ Grav.\  {\bf 34},  235006 (2017); 
  S.~Saini and P.~Singh,  Generic absence of strong singularities in loop quantum Bianchi-IX spacetimes, arXiv:1712.09474.
  
  \bibitem{cmb}  A.~Ashtekar and A.~Barrau,
  Class.\ Quant.\ Grav.\  {\bf 32},  234001 (2015); L.~C.~Gomar, M.~Martin-Benito and G.~A.~M.~Marug\'{a}n,
  JCAP {\bf 1506}, 045 (2015); I.~Agullo and P.~Singh,
 Loop Quantum Cosmology, in {\it{Loop
Quantum Gravity: The First 30 Years}}, Eds: A. Ashtekar, J. Pullin, World Scientific (2017) 
  arXiv:1612.01236; B.~Elizaga Navascu\'{e}s, M.~Martin-Benito and G.~A.~Mena Marug\'{a}n,
  Int.\ J.\ Mod.\ Phys.\ D{\bf 25}, 1642007 (2016);  T.~Zhu, A.~Wang, K.~Kirsten, G.~Cleaver and Q.~Sheng,
  Phys.\ Lett.\ B {\bf 773}, 196 (2017); E.~Wilson-Ewing,
  Comptes Rendus Physique {\bf 18}, 207 (2017);  and K. Kleidis, V.K. Oikonomou,  arXiv:1801.02578. 

  
  \bibitem{engle} J.~Engle,
  Class.\ Quant.\ Grav.\  {\bf 24}, 5777 (2007); J.~Brunnemann and C.~Fleischhack,
  ``On the configuration spaces of homogeneous loop quantum cosmology and loop quantum gravity,''
  arXiv:0709.1621; J.~Brunnemann and T.~A.~Koslowski,
  Class.\ Quant.\ Grav.\  {\bf 28}, 245014 (2011).

\bibitem{YDM09}  J. Yang, Y. Ding and Y. Ma, Phys. Lett. {\bf B}682 (2009) 1. 
  
  
\bibitem{thiemann} T. Thiemann, Class. Quant. Grav. {\bf 15}, 839 (1998); T. Thiemann,
Class. Quant. Grav. {\bf 15}, 875 (1998); K. Giesel, T. Thiemann, Class. Quant. Grav. {\bf 24}, 2465 (2007).  
  


\bibitem{martin} M. Bojowald, Class.Quant.Grav. 19, 2717 (2002).

\bibitem{cs08}  A.~Corichi and P.~Singh,
  Phys.\ Rev.\ D{\bf 78}, 024034 (2008).

\bibitem{abl} A. Ashtekar, M. Bojowald, and J. Lewandowski, Adv. Theor. Math. Phys. 7, 233 (2003).

\bibitem{DL17} A. Dapor and K. Liegener, Cosmological Effective Hamiltonian from full Loop Quantum Gravity Dynamics, arXiv:1706.09833.


\bibitem{states} T. Thiemann, O. Winkler, Class. Quant. Grav. {\bf 18} (2000) 2025; T. Thiemann, Class. Quant. Grav. {\bf 23} (2006) 2023.




\bibitem{kasner1}  B.~Gupt and P.~Singh,
  Phys.\ Rev.\ D{\bf 86}, 024034 (2012); A.~Corichi and P.~Singh, 
  Class.\ Quant.\ Grav.\  {\bf 33},  055006 (2016); A.~Corichi and E.~Montoya,
  Class.\ Quant.\ Grav.\  {\bf 34},  054001 (2017); E.~Wilson-Ewing,
  arXiv:1711.10943 [gr-qc].
  
  \bibitem{djs}  N.~Dadhich, A.~Joe and P.~Singh,
  Class.\ Quant.\ Grav.\  {\bf 32}, 185006 (2015).

\bibitem{adlp} M.~Assanioussi, A.~Dapor, K.~Liegener and T.~Pawlowski, Emergent de Sitter epoch of the quantum Cosmos,
  arXiv:1801.00768.
  
  

\bibitem{ps06} P.~Singh,
  Phys.\ Rev.\ D{\bf 73}, 063508 (2006).

\bibitem{Mei04}  K.A. Meissner, Class. Quantum Grav. {\bf 21} (2004) 5245.


 \bibitem{singh-soni} P. Singh, S. K. Soni,  Class. Quant. Grav. 33 (2016) 125001.

 \bibitem{CL04} S. M. Carroll and E. A. Lim, Phys. Rev. {\bf D}70, 123525 (2004).
 
 
 \bibitem{COb} Y. I. Izotov, T. X. Thuan and N. G. Guseva, Mon. Not. Roy. Astron. Soc. {\bf 445} (2014) 778;
 E. Aver, K. A. Olive and E. D. Skillman, JCAP {\bf 07},  011 (2015);
C. Patrignani et al. [Particle Data Group], Chin. Phys. {\bf C}40, 100001 (2016).


\bibitem{numerics} P.~Diener, B.~Gupt and P.~Singh,
  Class.\ Quant.\ Grav.\  {\bf 31}, 105015 (2014); P.~Diener, B.~Gupt, M.~Megevand and P.~Singh,
  Class.\ Quant.\ Grav.\  {\bf 31}, 165006 (2014); P.~Diener, A.~Joe, M.~Megevand and P.~Singh,
  Class.\ Quant.\ Grav.\  {\bf 34},  094004 (2017).

  
 \bibitem{DB} D. Baumann, TASI Lectures on Inflation, 
\href{http://arxiv.org/abs/0907.5424}{arXiv:0907.5424}.


\bibitem{AM15}
I. Agullo and N. A. Morris, 
\href{http://dx.doi.org/10.1103/PhysRevD.92.124040}{Phys. Rev. D{\bf 92}, 124040 (2015)}.


 
 \bibitem{Planck2015} P. Collaboration et al., Planck 2015. XX. Constraints on inflation, arXiv:1502.02114 [astro-ph].

 \bibitem{Zhu1}
T. Zhu, A. Wang, K. Kirsten, G. Cleaver,  and Q. Sheng,
\href{http://dx.doi.org/10.1103/PhysRevD.93.123525}{Phys. Rev. D{\bf 93}, 123525 (2016)}.


 \bibitem{Zhu2}  T. Zhu, A. Wang, K. Kirsten, G. Cleaver,  and Q. Sheng,
 Phys. Rev. {\bf D}96, 083520 (2017).  
 
 
  \bibitem{Shahalam}  M. Shahalam, M. Sharma, Q. Wu, and A. Wang,  Phys. Rev. {\bf D}96, 123533 (2017).  

  \bibitem{Agullo}
I. Agullo, 
\href{\Doi/10.1103/PhysRevD.92.064038}{Phys. Rev. D{\bf 92}, 064038 (2015)}.


   
 \bibitem{Zhu3} T. Zhu, A. Wang, K. Kirsten, G. Cleaver,  and Q. Sheng,  Primordial non-Gaussianity and power asymmetry
  with quantum gravitational effects in loop quantum cosmology, Phys. Rev. D{\bf 97}, 043501 (2018). 
  
 



\end{thebibliography}
\end{document}